\documentclass{article}

\usepackage{PRIMEarxiv}

\usepackage[utf8]{inputenc}
\usepackage[T1]{fontenc}
\usepackage{hyperref}
\usepackage{url}
\usepackage{booktabs}
\usepackage{array}
\usepackage{amsfonts}
\usepackage{amsmath}
\usepackage{amssymb}
\usepackage{microtype}
\usepackage{fancyhdr}
\usepackage{graphicx}
\usepackage{multirow}

\usepackage[numbers,sort&compress]{natbib}
\usepackage{float}
\usepackage{caption}
\usepackage{placeins}
\usepackage{algorithm}
\usepackage{algpseudocode}

\newcommand{\rev}[1]{#1}
\newcommand{\litrev}[1]{#1}

\graphicspath{{figs/}}

\title{Machine learning--assisted calibration of Agent-based Models: surrogate-based optimization with Genetic Algorithm and Particle Swarm Optimization}

\author{
  Duguma Yeshitla Habtemariam \\
  Department of Industrial and Data Engineering \\
  Major in Industrial Data Science and Engineering \\
  Pukyong National University, Busan, Republic of Korea \\
  \texttt{duguma@pukyong.ac.kr} \\
   \And
  {Jihwan Lee
    \thanks{\textit{Corresponding author}}
    }\\
  Department of Industrial and Data Engineering \\
  Major in Industrial Data Science and Engineering \\
  Pukyong National University, Busan, Republic of Korea \\
  \texttt{jihwan@pknu.ac.kr} \\
}

\begin{document}
\maketitle

\begin{abstract}
\litrev{Calibrating an agent-based model (ABM) is difficult because its objective landscape is stochastic and rugged, and can be evaluated only through costly black-box simulations. This study adapts \emph{inner-loop surrogate-assisted evolutionary computation (SAEC)} to ABM calibration by embedding a machine-learning surrogate within genetic algorithm (GA) and particle swarm optimisation (PSO). At each iteration, the surrogate screens the candidates and the simulator validates only the top \mbox{$50\%$}, reducing simulation demand while correcting surrogate errors. We evaluate a full factorial of \mbox{$48$} configurations combining two optimisers, five surrogates, and four calibration objectives on two contrasting ABMs, the Brock--Hommes asset-pricing model and the Island growth model. Parameter recovery is measured against pseudo-true values. Relative to the strongest pure-optimiser baseline, the best ML-assisted configurations reduce RMSE by \mbox{$20.0\%$} on Brock--Hommes and \mbox{$63.8\%$} on Island, while reducing computation time by \mbox{$32.1\%$} and \mbox{$61.1\%$}, respectively. ANOVA with Dunnett's post-hoc tests confirms significant time savings for every surrogate under both Brock--Hommes optimisers and under GA on Island. No surrogate differs significantly from the pure-optimiser baseline in parameter-recovery accuracy on Island; the reported accuracy gains are therefore best-configuration outcomes rather than average effects resolved at this sample size. The best surrogate--optimiser--objective combination changes with ABM complexity, indicating that surrogate-assisted calibration depends on the model and does not admit a universal default recipe.}
\end{abstract}


\section{Introduction}\label{sec:introduction}

Agent-based models (ABMs) are now indispensable wherever heterogeneity and out-of-equilibrium dynamics defeat analytical treatment, including computational economics and finance \cite{Platt2020,Lamperti2018}, epidemiology \cite{Michels2022}, and ecology \cite{Luke2019,An2012}. Their scientific use, however, rests on a single bottleneck, calibration to empirical data, that the modelling paradigm itself makes hard to perform. Calibration is the inverse problem of recovering parameter values under which the simulator's output reproduces a given empirical reference, naturally framed as the minimisation of a discrepancy measure, the calibration \emph{objective}, over the parameter space.

Two structural features make this minimisation difficult. First, the simulator is a non-analytic black box: gradients are unavailable, function values are stochastic across runs, and the objective landscape is rugged and multimodal even at moderate dimensionality. Second, each ABM run may consume seconds to hours of wall-clock time and calibration requires numerous simulation runs across the parameter space \cite{DAuria2020,Habtemariam2025}, so practical calibration budgets are tightly constrained. Gradient-free, population-based metaheuristics, most prominently the \emph{genetic algorithm} (GA) and \emph{particle swarm optimisation} (PSO) \cite{White2022,Michels2022,Alahmadi2024}, have become the dominant search choice in this regime, but each generation requires one ABM run per candidate, so a single calibration consumes a population-times-generations product of simulator calls, multiplied further by the replications needed to average over ABM stochasticity, a cost the practical budget cannot absorb. The standard remedy is to replace most of these evaluations with predictions from a much cheaper \emph{surrogate model} fitted on parameter--objective pairs accumulated during the search; yet a search driven solely by the surrogate has long been observed to converge to false optima introduced by approximation error rather than to the true objective \cite{Jin2000,Jin2011}.

\emph{Surrogate-assisted evolutionary computation} (SAEC) \cite{Jin2005,Jin2011} is the principled fix to this failure mode: the surrogate and the simulator are interleaved within a single search loop, a configuration we refer to throughout as \emph{inner-loop SAEC}, so that each generation the surrogate cheaply ranks all candidates, the simulator is queried only on the top-ranked fraction, and the resulting (parameter, objective) pairs are appended to the training set to retrain the surrogate before the next round. The simulator thereby acts as a corrective oracle against surrogate-induced false optima, while the surrogate compresses per-generation simulation cost. This arrangement, in which the surrogate ranks all candidates each generation and the simulator validates the top fraction, is the ``best-strategy'' model-management pattern of surrogate-assisted evolutionary computation \cite{Jin2005,Jin2011,Branke2005}. \textbf{Despite its clean structural fit with the noisy, expensive evaluations characteristic of ABMs, the pattern has seen little uptake in ABM calibration.} ABM studies that invoke surrogates typically do so in narrower forms: fitting them offline on a fixed batch of runs and then optimising over the frozen response surface \cite{Lamperti2018,Robertson2025}, using them exclusively for post-hoc sensitivity analysis \cite{Angione2022}, or training a learned emulator of the agent dynamics on empirical agent-level data and embedding it inside a single optimiser in place of the ABM \cite{PSODeepLearning2024}. The active, online role of the surrogate as a screening oracle inside the iterative search loop (and, more importantly, a systematic comparison of how its choice interacts with the optimiser and the calibration objective) has yet to be operationalised in the ABM-calibration literature.

\textbf{The present paper makes precisely this transfer.} We embed an ML surrogate as an active screening oracle inside the iterative search loop of both GA and PSO for ABM calibration, and evaluate the resulting framework through a full factorial that varies surrogate, optimiser, and calibration objective jointly within the loop, with the unsurrogated metaheuristics retained as baselines. The surrogate dimension spans linear, kernel-based, tree-based, and neural-network architectures; the objective dimension spans moment-, information-, divergence-, and distribution-based discrepancy measures. The framework is tested on two ABMs that have served as standard benchmarks for surrogate-based calibration in prior work \cite{Lamperti2018,Platt2020}, the Brock--Hommes asset-pricing model \cite{BrockHommes1998} and the Island growth model \cite{FagioloDosi2003}, chosen for their contrasting dimensionality and emergent dynamics. Recovered parameters are judged against a \emph{pseudo-true} ground truth, complemented by analysis-of-variance against the baselines and Shapley-value attribution on the calibrated surrogate.

Across both ABMs, inner-loop SAEC reduces parameter-recovery RMSE by \rev{20.0\%} on the Brock--Hommes model and \rev{63.8\%} on the Island model relative to the strongest unsurrogated baseline, while completing \rev{32.1\%} and \rev{61.1\%} faster, respectively. \rev{These claims are tested with one-way ANOVA and Dunnett's post-hoc tests, which compare each surrogate against the unsurrogated optimiser across all four objectives and all five runs. The speed advantage is unambiguous: on Brock--Hommes every surrogate is significantly faster than the pure optimiser under both GA and PSO, and on Island the same holds under GA. Accuracy behaves differently. On Island no surrogate shifts recovery error significantly in either direction; most of the differences point the right way, but the spread between independent runs is wider than the gaps between surrogates, so at \mbox{$n = 20$} observations per group the tests cannot separate them. On Brock--Hommes the only significant accuracy effects are adverse ones, from the two simplest surrogates (linear regression and SVM) under GA, while the more expressive surrogates are statistically indistinguishable from the baseline. The RMSE reductions quoted above therefore describe the best configuration found rather than a surrogate effect separated from run-to-run variation.} The calibration objective emerges as the dominant factor in fit quality, and the optimal surrogate--optimiser--objective combination is model-dependent, evidence that argues against any single default calibration recipe.

The contributions are threefold. \emph{Methodologically}, we adapt the established inner-loop SAEC pattern \cite{Jin2005,Jin2011} to ABM calibration: an ML surrogate is operated as an active screening oracle inside the iterative search loop of GA and PSO, within a unified framework that accommodates heterogeneous surrogate architectures. \emph{Empirically}, a $48$-configuration factorial across two ABMs of contrasting complexity provides systematic evidence on how surrogate, optimiser, and calibration objective interact in the inner loop, a comparison absent from the existing ABM-calibration literature, and shows that the dominant surrogate--optimiser--objective configuration is model-dependent. \emph{Analytically}, the framework is paired with one-way ANOVA + Dunnett's tests and Shapley-value attribution to surface signed parameter sensitivities at no additional simulation cost, turning the calibration output into a calibrated \emph{and} interpretable model.

The remainder of the paper is organised as follows: Section~\ref{sec:literature} surveys related work, Section~\ref{sec:methodology} develops the ML-assisted methodology, Section~\ref{sec:experiments} reports the experimental design, Section~\ref{sec:results} presents the results including the ANOVA and SHAP analyses, and Section~\ref{sec:conclusion} concludes.

\section{Literature review}\label{sec:literature}

\subsection{ABM calibration: state of the art}\label{sec:lit-pillar-abm}

\litrev{ABM calibration involves three main design choices:} the search algorithm that explores the parameter space, the surrogate that (when used) replaces a fraction of expensive ABM evaluations, and the goodness-of-fit measure that defines the objective being optimised. Existing studies typically vary one or two of these dimensions while holding the others fixed, \litrev{so how all three interact within a unified scheme remains largely unaddressed.}

\paragraph{Population-based search:} \litrev{ABM calibration is fundamentally a search over a parameter space that is typically high-dimensional, non-smooth, and noisy. These conditions rule out gradient-based methods.} Genetic algorithms (GA) and particle swarm optimisation (PSO) are the two dominant population-based metaheuristics in this regime \cite{White2022,Michels2022}. \litrev{Both are derivative-free and tolerate multi-modal, noisy landscapes, but they balance broad exploration and convergence speed differently} \cite{Alahmadi2024}. White et al.\ \cite{White2022} demonstrate effective GA-based search of high-dimensional ABM parameter spaces, \litrev{while} Michels et al.\ \cite{Michels2022} report that PSO outperforms Monte Carlo search for a spatially explicit influenza ABM, attributing the gain to PSO's robustness on noisy, non-convex performance surfaces. Both methods share a structural drawback: every candidate generated in every iteration requires a full ABM evaluation. \litrev{This computational bottleneck motivates the surrogate-based approaches reviewed below.}

\paragraph{Surrogate-based methods in ABM calibration:} To alleviate this computational burden, the ABM community has increasingly turned to \emph{surrogate models} that approximate the parameter-to-objective mapping from a limited number of full ABM evaluations \cite{Sivakumar2022,PlatasLopez2025,Ionescu2024,Turgut2023}. Turgut and Bozdag~\cite{Turgut2023} \litrev{provide a systematic framework for the broader relationship between ML and ABMs. They map how supervised learning can model agent behaviours directly and use a real-world case study to evaluate the benefits and drawbacks of data-driven ABMs. The setting considered here is one distinct form of this broader relationship: ML surrogates replace expensive simulator evaluations during parameter search. The authors identify this as a promising avenue but do not operationalise it.} Three modes of surrogate use have emerged. The dominant mode is \emph{offline} fitting, in which the surrogate is trained on a precomputed batch of ABM evaluations and used to select further candidates. Lamperti et al.\ \cite{Lamperti2018} introduce a non-parametric ML surrogate for ABM calibration. \litrev{They evaluate an initial parameter set in the ABM, fit a surrogate to the resulting objective values, and use its predictions to select further candidates. The surrogate is retrained between rounds until a fixed evaluation budget is exhausted.} Although iterative across rounds, this scheme optimises over a frozen response surface within each round rather than continuously interleaving the surrogate and the simulator. \litrev{Similarly,} Robertson et al.\ \cite{Robertson2025} fit Random Forest surrogates globally to approximate likelihood surfaces, again outside any inner search loop. A second mode treats the surrogate as a \emph{post-hoc diagnostic}: Angione et al.\ \cite{Angione2022} train surrogates retrospectively on ABM input--output pairs to assess parameter sensitivity and benchmark architectures. \litrev{They find that neural networks and gradient-boosted trees outperform Gaussian processes when training sets are large.} Subsequent work in the same vein \cite{Perumal2020,Llacay2025,Zhang2020} further benchmarks tree-based, neural, and kernel-based architectures as the principal contenders for ABM-side surrogation. A related extension targets the \emph{sampling strategy} rather than the surrogate architecture. Edali and Y{\"u}cel~\cite{Edali2019} combine Random Forest metamodels with uncertainty-based sequential sampling to explore the \litrev{behaviour} space of ABMs. \litrev{They show that repeatedly targeting the most uncertain parameter regions produces significantly more accurate metamodels than random designs. This approach also naturally reveals tipping-point boundaries and counter-intuitive dynamics that random sampling misses.} A further extension of this post-hoc diagnostic mode combines surrogate fitting with interpretable machine learning. De Bosscher et al.\ \cite{DeBosscher2023} train Gaussian process, gradient-boosted, and random-forest surrogates on the Agent-based Airport Terminal Operations Model (AATOM) via active learning. \litrev{They then apply state-of-the-art explainable-AI methods to the fitted surrogates to link emergent phenomena to specific parameter interactions. These phenomena include knock-on effects on passenger discretionary spending and security-checkpoint saturation. Their results demonstrate that surrogates trained on ABM output can serve as interpretable proxies for understanding complex sociotechnical dynamics.} A third mode replaces the simulator entirely with a \emph{learned emulator of the agent dynamics}. Amartaivan et al.\ \cite{PSODeepLearning2024} train a TabNet model on real-world pedestrian trajectories to predict each agent's next-step velocity from local conditions and tunable parameters. \litrev{They then embed this emulator inside a PSO loop in place of the ABM itself, reserving the simulator for post-calibration visualisation rather than querying it during the search.} \litrev{This arrangement differs structurally from the inner-loop screening pattern reviewed below.} The surrogate substitutes for the simulator rather than ranking candidates against it, and \litrev{the ABM does not serve as a corrective oracle during calibration.} In none of these three ABM-calibration modes does the surrogate operate as an \emph{active}, online screening oracle within the iterative loop of an evolutionary search.

\paragraph{Goodness-of-fit measures:} \litrev{The third design choice, alongside the search algorithm and surrogate, is the calibration objective: the formal criterion used to compare simulator output with observed data.} The choice of metric is consequential, yet most ABM studies adopt a single objective without systematic comparison \cite{Platt2020}. The most direct evidence comes from Platt \cite{Platt2020}, whose multi-metric study across four benchmark ABMs shows that moment-based metrics (MSM) are tractable but sensitive to moment selection, while information-theoretic objectives (MIC, GSL-div) better capture sequential time-series structure. Beyond this systematic comparison, individual studies illustrate the variety of metrics \litrev{used} in practice: Lamperti et al.\ \cite{Lamperti2018}, whose surrogate scheme was described above, adopt the Kolmogorov--Smirnov (KS) statistic \cite{Massey1951} on simulated returns from the Brock--Hommes model; Fransen and Davydenko \cite{Fransen2021} minimise the sum of squared errors in a port-logistics ABM; and Robertson et al.\ \cite{Robertson2025} adopt a Bayesian approach for a large-scale urban contagion ABM, evaluating fit through posterior predictive diagnostics. Metric choice is, therefore, a substantive modelling decision rather than a neutral preprocessing step. \litrev{The evolutionary-computation paradigm reviewed next provides a framework that embeds the surrogate as an active, online oracle while remaining agnostic to the choice of metric.}

\subsection{Surrogate-assisted evolutionary computation}\label{sec:lit-pillar-saec}

In the evolutionary-computation literature, the same cost-versus-fidelity tension is addressed by a mature paradigm: \emph{surrogate-assisted evolutionary computation} (SAEC). \litrev{Here, a cheap learned approximator and the expensive true objective are interleaved within a single search loop.} \litrev{Because surrogate predictions inevitably contain approximation error, a search driven solely by the surrogate can converge to spurious optima introduced by the approximator rather than to the true objective} \cite{Jin2000}. The SAEC remedy is to retain the simulator as a \emph{corrective oracle}: the surrogate reduces per-generation cost while the simulator validates a controlled subset of candidates and supplies fresh training data back to the surrogate. \litrev{This paradigm has a long history in evolutionary computation, including surrogate-guided global-search procedures such as the radial-basis-function and Latin-hypercube scheme of Knysh and Korkolis} \cite{Knysh2016}, \litrev{but it has seen little uptake in ABM calibration.}

\paragraph{Model management strategies:} \litrev{Putting this idea into practice requires principled} \emph{model management}\litrev{: a framework for deciding when to use the surrogate and when to run the true evaluation} \cite{Jin2011}. Three strategies have been identified \cite{Jin2005,Jin2011}: \emph{individual-based} (the real function is applied to selected individuals per generation), \emph{generation-based} (generations alternate between surrogate and real evaluations), and \emph{population-based} (multiple sub-populations each guided by their own surrogate). Within individual-based management, the \emph{best strategy} ranks all offspring by the surrogate and re-evaluates only the top subset with the real function, \litrev{while assigning surrogate-predicted fitness to the rest.} The related \emph{pre-selection} variant screens a larger pool so that all parents entering the next generation carry true fitness values \cite{Jin2011}. The variant in which the surrogate is queried at every iteration to filter all candidates, which we term \emph{inner-loop SAEC}, falls within this individual-based, best-strategy family. \litrev{Recent work continues to extend this pattern across application domains.} Wang et al.\ \cite{WangCAL2017} apply it to expensive single-objective optimisation in engineering design, including a transonic airfoil shape problem. Liu et al.\ \cite{LiuSAMaOEA2026} apply it to expensive super-many-objective optimisation, \litrev{validating it on suitable benchmark suites and a water-resource design problem.} Mej\'ia-de-Dios et al.\ \cite{MejiadeDios2025} apply it to bilevel location-routing in a hierarchical supply chain, where the surrogate spares the upper level from the cost of repeatedly invoking the lower-level optimiser. Pineda et al.\ \cite{Pineda2025} apply it to multi-server queuing system design. \litrev{They embed a machine-learning surrogate inside an evolutionary search loop to reduce reliance on expensive discrete-event simulations and achieve solution quality comparable to full simulation optimisation at significantly lower computational cost. This pattern is structurally identical to the one transferred to ABM calibration in the present paper.} Chen et al.\ \cite{Chen2026} extend the pattern to multi-objective simulation-based optimisation of cold spray additive manufacturing parameters. \litrev{They replace expensive SPH simulations with an ensemble of surrogates and introduce a dedicated infill criterion to govern when the simulator is queried. Their study illustrates that principled model management generalises beyond single-objective search to high-cost, noisy simulation contexts.}

\paragraph{Robustness in stochastic settings:} The inner-loop arrangement is particularly valuable when the objective is stochastic. Because ABM losses are noisy across Monte Carlo replications, a surrogate trained on multiple $(\boldsymbol{\theta}, \mathcal{L}(\boldsymbol{\theta}))$ pairs averages over this variation and produces a smoother approximation of the conditional expectation $\mathbb{E}[\mathcal{L}(\boldsymbol{\theta}) \mid \boldsymbol{\theta}]$ \cite{Branke2005,Jin2011}. \litrev{This trade-off, some added approximation bias in exchange for sharply reduced variance, improves search outcomes on noisy, multi-modal landscapes.} \litrev{In practical terms, the surrogate in the inner-loop regime not only reduces computational cost but also denoises the objective. This effect does not arise from offline or post-hoc surrogate use.} This noise-reduction benefit has been demonstrated empirically in adjacent stochastic simulation domains, including queuing system design with stochastic demand variables \cite{Pineda2025}.

\subsection{Research gap}\label{sec:lit-gap}

\litrev{The two pillars reviewed above have yet to be integrated. The first is ABM calibration practice, with its three loosely connected design choices. The second is the SAEC paradigm, with its model-management strategies for embedding surrogates within iterative search loops.} Although Jin \cite{Jin2005,Jin2011} formalised inner-loop embedding as the principled model-management strategy long ago, the ABM calibration literature \litrev{has seen little uptake of this pattern. The literature itself has been criticised for compartmentalisation and reliance on highly simplified test models} \cite{Platt2020}. \litrev{Existing ABM-surrogate work fits surrogates offline, deploys them post hoc, or couples a single surrogate with a single optimiser on a single ABM.} The gap, then, is one of \emph{domain transfer and systematic comparison} rather than of methodological invention. \litrev{What remains under-explored in the ABM setting is the active, online role of the surrogate within the iterative search loop, together with a systematic comparison of how the surrogate, optimiser, and goodness-of-fit measure interact in that role.} Even the most interpretability-oriented surrogate work, which couples surrogates with explainable AI to attribute emergent ABM phenomena to parameter interactions \cite{DeBosscher2023}, operates outside any calibration search loop and does not address parameter recovery. The framework developed in the next section addresses this gap.

\section{Methodology}\label{sec:methodology}

\subsection{Framework}\label{sec:framework}

\litrev{The inner-loop SAEC framework addresses three main obstacles in ABM calibration:}

\begin{itemize}
  \item \textbf{Stochastic, gradient-free objective:} \litrev{Addressed through population-based search using GA or PSO} \cite{Alahmadi2024,White2022,Michels2022}.
  \item \textbf{Per-iteration ABM cost:} The surrogate ranks all candidates each iteration, \litrev{and only the top} 50\% \litrev{are sent to the ABM for true evaluation.}
  \item \textbf{Surrogate-induced false optima:} \litrev{The simulator acts as a corrective oracle, while the surrogate is retrained on each new batch of true evaluations} \cite{Jin2000,Jin2011}.
\end{itemize}

\begin{figure}[p]
  \centering
  \includegraphics[width=0.9\textwidth]{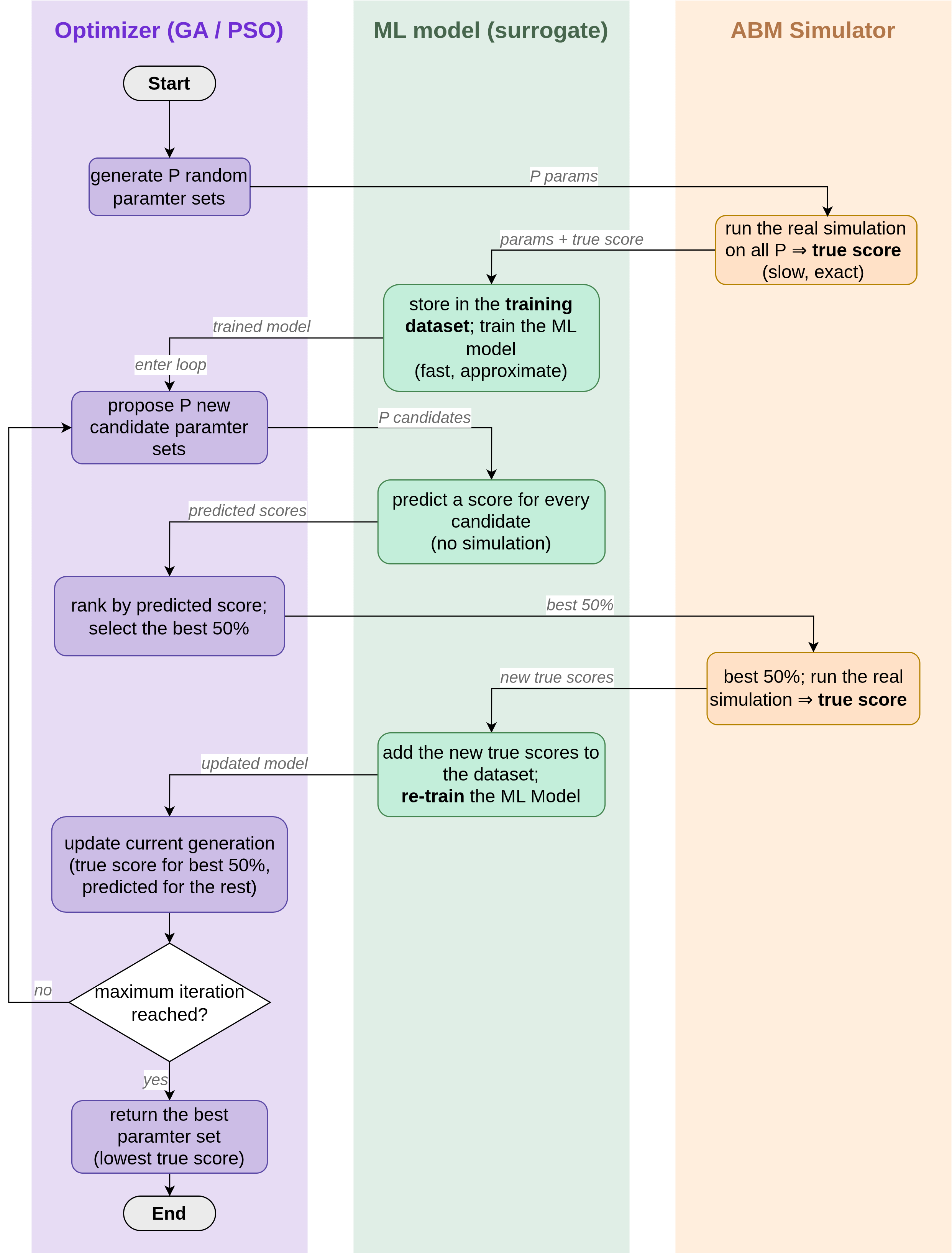}
  \caption{\protect\litrev{Inner-loop SAEC workflow across the optimiser, ML surrogate, and ABM simulator. The simulator first evaluates} $P$ \protect\litrev{random parameter sets to train the surrogate. At each iteration, the optimiser proposes} $P$ \protect\litrev{candidates, the surrogate ranks them, and the simulator evaluates the best} $50\%$\protect\litrev{. The resulting true scores retrain the surrogate and update the population, while the remaining candidates retain predicted scores. The loop continues until the stopping criterion is met and returns the parameter set with the lowest estimated error.}}
  \label{fig:saec-framework}
\end{figure}

\litrev{The framework brings five components together in a single iterative loop:} a population-based optimiser (GA or PSO), an ML surrogate $f$ that approximates the calibration objective, a calibration objective $\mathcal{L}(\boldsymbol{\theta};\mathbf{y})$ that defines model--data fit, a screening fraction (50\%) that splits surrogate-ranked candidates between true and predicted evaluation, and a retraining cadence that refreshes $f$ after every batch of ABM evaluations. \litrev{The loop continues under a fixed simulation budget of} $N_{\text{ABM}}$. Figure~\ref{fig:saec-framework} summarises the overall structure. \litrev{An initial ABM-only sample seeds the surrogate. The loop then alternates among candidate generation, surrogate-based ranking,} top-$50\%$ \litrev{ABM validation, and surrogate retraining until the budget is exhausted.} Section~\ref{sec:ml-algorithm} develops the algorithm; Section~\ref{sec:surrogates} specifies the surrogate ensemble; Section~\ref{sec:objectives} specifies the four calibration objectives evaluated.

\subsection{Inner-loop SAEC algorithm}\label{sec:ml-algorithm}

\litrev{The algorithm uses the surrogate as an active screening oracle within the iterative search loop of a population-based optimiser. The process has two phases.} \emph{Phase 1: initial exploration.} \litrev{First,} $P$ \litrev{candidate parameter vectors are sampled within the bounds and evaluated with the ABM. These evaluations form the initial training set} $(\mathcal{X}_{\text{train}}, \mathcal{Y}_{\text{train}})$\litrev{, which is used to train the surrogate} $f$\litrev{. The optimiser is then initialised with these} $P$ \litrev{evaluated individuals.} \emph{Phase 2: screened iteration.} \litrev{At each iteration, the optimiser proposes a new batch of} $P$ \litrev{candidates. The surrogate predicts} $\widehat{z} = f(\boldsymbol{\theta})$ \litrev{for every candidate. The top} 50\% \litrev{based on} $\widehat{z}$ \litrev{are sent to the ABM for true evaluation; the remaining candidates are not evaluated by the ABM in that iteration. The new ABM evaluations are added to} $(\mathcal{X}_{\text{train}}, \mathcal{Y}_{\text{train}})$\litrev{, and the surrogate is retrained.} The loop terminates when $N_{\text{ABM}}$ is exhausted, returning the individual with the best true ABM fitness. This top-fraction screening follows the \emph{best strategy} model-management approach of \cite{Jin2005,Jin2011}: the simulator validates only the most promising candidates while the surrogate compresses per-iteration cost. \litrev{Algorithms~}\ref{alg:ga-ml} \litrev{and~}\ref{alg:pso-ml} \litrev{apply this loop to GA and PSO, respectively. Their mechanisms are summarised next.}

\paragraph{GA instantiation:}\label{sec:ga} A genetic algorithm maintains a population of real-coded parameter vectors. It improves the population through tournament selection, real-coded crossover, and Gaussian mutation. \litrev{Because it is gradient-free, the algorithm tolerates noisy, multimodal, and non-smooth landscapes} \cite{Holland1975,Alahmadi2024,White2022}. \litrev{Within the inner-loop SAEC scheme, the candidates at each iteration are the} $P$ \litrev{offspring generated from the current ranked population by the GA's variation operators. The screening rule and retraining cadence from Phase 2 then apply without change.} A standard GA requires $P \times G$ ABM evaluations over $G$ generations, whereas surrogate screening reduces the total to $P + 0.5P(G-1)$; for the default settings $P = G = 50$ used here, $1{,}275$ versus $2{,}500$ evaluations, a $\approx49\%$ reduction.

\begin{algorithm}
\caption{GA + ML surrogate integration}
\label{alg:ga-ml}
\begin{algorithmic}[1]
\Require Target data $\mathbf{y}$; ABM simulator; objective $\mathcal{L}$; ML surrogate $f$; simulation budget $N_{\text{ABM}}$; population size $P$; top fraction (50\%).
\Ensure Calibrated parameter vector $\widehat{\boldsymbol{\theta}}$.
\Statex
\State \textbf{Initial exploration (ABM only)}
\State Sample $P$ candidate parameter vectors within bounds.
\State Evaluate each candidate with the ABM to obtain objective values.
\State Construct initial training data $(\mathcal{X}_{\text{train}}, \mathcal{Y}_{\text{train}})$ from evaluated candidates.
\State Train surrogate $f$ on $(\mathcal{X}_{\text{train}}, \mathcal{Y}_{\text{train}})$.
\State Initialise the GA population with these $P$ evaluated individuals.
\Statex
\State \textbf{Evolution with surrogate screening}
\While{$n_{\text{ABM}} < N_{\text{ABM}}$}
  \State Rank the current population by fitness
  \State Generate $P$ offspring via selection, crossover, and mutation.
  \State $\triangleright$ \emph{Approximate each offspring's fitness via the surrogate rather than the ABM.}
  \ForAll{offspring $\boldsymbol{\theta}$}
    \State $\widehat{z} \gets f(\boldsymbol{\theta})$ \Comment{Surrogate-predicted fitness}
  \EndFor
  \State Rank offspring by $\widehat{z}$ and select the top 50\%.
  \State Evaluate the selected offspring with the ABM; update $n_{\text{ABM}}$.
  \State Augment $(\mathcal{X}_{\text{train}}, \mathcal{Y}_{\text{train}})$ with new ABM evaluations.
  \State Retrain $f$ on the updated training set.
  \State Update the GA population using offspring with true fitness where available.
\EndWhile
\Statex
\State \textbf{Return best solution}
\State $\widehat{\boldsymbol{\theta}} \gets$ best individual in the population by true ABM fitness.
\State \Return $\widehat{\boldsymbol{\theta}}$
\end{algorithmic}
\end{algorithm}

\paragraph{PSO instantiation:}\label{sec:pso} A particle swarm maintains $N$ particles, each with position $\mathbf{x}_i^t$ and velocity $\mathbf{v}_i^t$ updated from the particle's personal best $\mathbf{p}_i$ and the swarm's global best $\mathbf{p}_g$ \cite{Alahmadi2024,Michels2022}. The standard inertia-form update is
\begin{align}
\mathbf{v}_i^{t+1} &\leftarrow \omega\, \mathbf{v}_i^{t} + \mathbf{U}(0,\phi_1) \odot (\mathbf{p}_i - \mathbf{x}_i^{t}) + \mathbf{U}(0,\phi_2) \odot (\mathbf{p}_g - \mathbf{x}_i^{t}), \label{eq:pso-basic-v}\\
\mathbf{x}_i^{t+1} &\leftarrow \mathbf{x}_i^{t} + \mathbf{v}_i^{t+1}, \label{eq:pso-basic-x}
\end{align}
where $\omega$ is the inertia weight, $\phi_1$ and $\phi_2$ scale the pull toward personal and global bests, $\mathbf{U}(0,\phi)$ draws each component uniformly from $[0,\phi]$, and $\odot$ is the elementwise product. \litrev{Within the inner-loop SAEC scheme, the candidates at each iteration are the} $N$ \litrev{particle positions produced by the velocity update. The surrogate predictions are used only to rank the particles and select the top} $50\%$ \litrev{for ABM evaluation. These true evaluations then update the personal and global bests} $\mathbf{p}_i$ \litrev{and} $\mathbf{p}_g$\litrev{. The unselected particles are not re-evaluated and retain their existing personal and global bests.} A standard PSO requires $N \times T$ ABM evaluations over $T$ iterations, whereas surrogate screening reduces the total to $N + 0.5N(T-1)$; for the default settings $N = T = 50$ used here, $1{,}275$ versus $2{,}500$ evaluations, a $\approx49\%$ reduction.

\begin{algorithm}
\caption{PSO + ML surrogate integration}
\label{alg:pso-ml}
\begin{algorithmic}[1]
\Require Target data $\mathbf{y}$; ABM simulator; objective $\mathcal{L}$; ML surrogate $\mathcal{M}$; simulation budget $N_{\text{ABM}}$; swarm size $N$; top fraction (50\%).
\Ensure Calibrated parameter vector $\boldsymbol{\theta}^*$.
\Statex
\State \textbf{Initial exploration (ABM only)}
\State Sample $N$ initial particle positions within parameter bounds.
\State Evaluate each position with the ABM to obtain objective values.
\State Build initial training data $(\mathcal{X}_{\text{train}}, \mathcal{Y}_{\text{train}})$ and train $\mathcal{M}$.
\State Initialise swarm: positions $\mathbf{x}_i$, velocities $\mathbf{v}_i$, personal bests $\mathbf{p}_i$, and global best $\mathbf{g}$.
\Statex
\State \textbf{PSO iteration with surrogate screening}
\While{ABM evaluation count $< N_{\text{ABM}}$}
  \State Update velocities and positions for all particles (standard PSO update using inertia, personal best, and global best).
  \State $\triangleright$ \emph{Approximate each particle's fitness via the surrogate rather than the ABM.}
  \ForAll{particles $i = 1,\dots,N$}
    \State $\widehat{f}_i \gets \mathcal{M}(\mathbf{x}_i)$ \Comment{Surrogate-predicted fitness}
  \EndFor
  \State Rank particles by $\widehat{f}_i$ and select the top 50\%.
  \State Evaluate selected particles with the ABM; update personal bests $\mathbf{p}_i$ and global best $\mathbf{g}$; update ABM evaluation count.
  \State Append new ABM evaluations to $(\mathcal{X}_{\text{train}}, \mathcal{Y}_{\text{train}})$ and retrain $\mathcal{M}$.
\EndWhile
\Statex
\State \textbf{Return best solution}
\State $\boldsymbol{\theta}^* \gets \mathbf{g}$.
\State \Return $\boldsymbol{\theta}^*$
\end{algorithmic}
\end{algorithm}

\subsection{Surrogate ensemble}\label{sec:surrogates}

The surrogate $f$ approximates the parameter--objective mapping $\boldsymbol{\theta} \mapsto \mathcal{L}(\boldsymbol{\theta};\mathbf{y})$ from the running training set $(\mathcal{X}_{\text{train}}, \mathcal{Y}_{\text{train}})$. \litrev{This training set grows across iterations, allowing the surrogate to estimate the fitness of any new} $\boldsymbol{\theta}$ \litrev{at negligible cost.} We evaluate five architectures spanning linear, kernel-based, tree-based, and neural function classes:

\begin{enumerate}
  \item \textbf{LR}: linear regression fitted by ordinary least squares; linear baseline
  \item \textbf{SVM}: support vector regression with RBF kernel
  \item \textbf{XGBoost}: gradient-boosted decision trees \cite{Perumal2020,Angione2022}
  \item \textbf{Random forest}: bagged decision-tree ensemble \cite{Robertson2025}
  \item \textbf{ANN}: feedforward, two hidden layers, ReLU activations, Adam optimiser \cite{Angione2022}
\end{enumerate}

\litrev{The architectures were chosen because their training cost remains well below the cost of the ABM evaluations they replace. This requirement defines the framework's operating envelope and excludes architectures} (e.g.\ large deep networks) \litrev{whose per-iteration retraining would reduce the efficiency gain.}

\subsection{Calibration objectives}\label{sec:objectives}

\litrev{The calibration objective} $\mathcal{L}(\boldsymbol{\theta};\mathbf{y})$ \litrev{serves two roles in the inner-loop SAEC scheme. It provides the true fitness signal whenever the ABM is queried and is also the target that the surrogate} $f$ \litrev{is trained to predict. The choice of} $\mathcal{L}$ \litrev{therefore shapes both what the surrogate learns and the fitness landscape that the optimiser navigates.} We evaluate four objectives that span moment-, information-, divergence-, and distribution-based discrepancy measures.

\paragraph{Method of simulated moments (MSM):}\label{sec:obj-msm}
MSM compares descriptive moments of the simulated and empirical series and minimises a weighted sum of their squared differences \cite{Alahmadi2024}. A small set of statistics characterises the time series: variance, kurtosis, and a few autocorrelation values capture, respectively, spread, tail behaviour, and temporal dependence. A well-calibrated parameter set produces simulated output whose moments match the empirical ones closely; mismatches produce a large MSM score, and the calibration algorithm searches for the parameter values that minimise it. We use the seven-moment vector (variance, kurtosis, and five autocorrelations of the raw, absolute, and squared series) and weight squared differences by the inverse Newey--West covariance of the empirical moments (see Appendix~\ref{app:msm}).

\paragraph{Markov Information Criteria (MIC):}\label{sec:obj-mic}
\litrev{MIC treats calibration as a prediction task. A compact Markov model trained on the simulated series is used to predict the empirical series. The difference between its loss and that of a Markov model trained directly on empirical data measures the mismatch in temporal structure} \cite{Barde2017,Alahmadi2024}. A small gap indicates that simulated dynamics share the same sequential structure as the data; a large gap signals temporal mismatch. Both series are first discretised into binary symbol sequences, making MIC applicable even when the data are non-Gaussian or exhibit complex non-linear patterns \cite{Platt2020} (see Appendix~\ref{app:mic}).

\paragraph{Generalised subtracted $L$-divergence (GSL-div):}\label{sec:obj-gsl}
GSL-div compares the temporal patterns of the simulated and empirical series without choosing a fixed set of summary statistics \cite{Lamperti2016,Alahmadi2024}. \litrev{First, both series are represented using an alphabet of} $b$ \litrev{symbols. Short patterns of length} $l$ \litrev{are then counted at multiple window lengths. At each scale, the resulting pattern-frequency distributions are compared using a symmetric} $L$\litrev{-divergence. Finally, the contributions across scales are combined into a single bounded score with a small-sample bias correction.} A lower score indicates closer agreement of pattern frequencies across time scales, making GSL-div particularly sensitive to the dynamic and sequential properties of complex systems (see Appendix~\ref{app:gsl}).

\paragraph{Kolmogorov--Smirnov test (KS):}\label{sec:obj-ks}
The KS statistic is the maximum absolute difference between two cumulative distribution functions and is non-parametric in both shape and family \cite{Massey1951}. Applied to ABM calibration, it compares the empirical CDF of simulated output with that of the target; a smaller statistic indicates closer agreement of the full distributional shape rather than selected moments. We use a weighted combination of KS terms over the raw, absolute, and squared series; weights $w_X$, $w_{|X|}$, $w_{X^2}$ are reported in Table~\ref{tbl:objectives-hp}.

\section{Experimental setup}\label{sec:experiments}

\litrev{This section evaluates the inner-loop SAEC framework introduced in} Section~\ref{sec:methodology}. \litrev{The evaluation follows a full factorial design that combines} two optimisers (GA, PSO), five surrogates (LR, SVM, XGBoost, RF, ANN), and four calibration objectives (MSM, MIC, GSL-div, KS). \litrev{This gives} $2 \times 4 + 2 \times 5 \times 4 = 48$ \litrev{configurations: 8 pure-optimiser baselines and 40 ML-assisted variants. All configurations use a common simulation budget and the same top-}50\% \litrev{screening policy. Each configuration is tested on two benchmark ABMs with different dimensionalities.} Section~\ref{sec:benchmark-models} introduces the benchmarks; Section~\ref{sec:expt-calibration} describes the pseudo-true calibration protocol; Section~\ref{sec:expt-config} lists the experimental configuration; Section~\ref{sec:expt-sensitivity} introduces the post-experiment SHAP analysis.

\subsection{Benchmark models}\label{sec:benchmark-models}

\paragraph{Brock--Hommes (BH):}\label{sec:bh-model}
The Brock--Hommes model \cite{BrockHommes1998} is a stylised asset-pricing ABM in which a population of traders \litrev{invests} in either a risk-free bond or a risky asset. Traders are heterogeneous: each follows one of $H$ trading rules. Some are \emph{fundamentalists} who expect prices to revert toward a fair value; others are \emph{trend-followers} (chartists) who extrapolate recent movements. Each rule predicts next-period price deviations using a linear belief rule $g_h x + b_h$, where $g_h$ is the trend-following coefficient and $b_h$ the bias. \litrev{At each step, the market price is determined by aggregating traders' demands. Traders then switch strategies using softmax weights based on recent profitability} \cite{Platt2020}.

In our experiments, the model has $H = 4$ strategy types: strategy~1 is a pure fundamentalist ($g_1 = b_1 = 0$); strategies~2 and~3 carry the four free calibration parameters $\boldsymbol{\theta} = (g_2, b_2, g_3, b_3)$; strategy~4 is neutral ($g_4 = b_4 = 0$). Global parameters $\beta$ (intensity of choice), $\sigma$ (noise), and $r$ (risk-free return) are held fixed at the values in Table~\ref{tbl:bh-params}. Series length is $T = 1000$. See Appendix~\ref{app:bh} for the full derivation.

\begin{table}[!t]
\centering
\caption{BH model parameters for the pseudo-true data generation experiment.}
\label{tbl:bh-params}
\begin{tabular}{@{}llccc@{}}
\toprule
Type & Parameter & Symbol & Pseudo-true value & Search bounds \\
\midrule
\multirow{3}{*}{Fixed global} & Risk-free return & $r$ & 1.05 & --- \\
 & Intensity of choice & $\beta$ & 1.0 & --- \\
 & Noise std dev & $\sigma$ & 0.05 & --- \\
\midrule
\multirow{4}{*}{Fixed strategy} & Strategy 1 trend & $g_1$ & 0 & --- \\
 & Strategy 1 bias & $b_1$ & 0 & --- \\
 & Strategy 4 trend & $g_4$ & 0 & --- \\
 & Strategy 4 bias & $b_4$ & 0 & --- \\
\midrule
\multirow{4}{*}{Free (calibrated)} & Strategy 2 trend & $g_2$ & 0.6 & $[-2,\,2]$ \\
 & Strategy 2 bias & $b_2$ & 0.2 & $[-2,\,2]$ \\
 & Strategy 3 trend & $g_3$ & 0.7 & $[-2,\,2]$ \\
 & Strategy 3 bias & $b_3$ & $-0.2$ & $[-2,\,2]$ \\
\midrule
\multirow{3}{*}{Simulation} & Series length & $T$ & 1000 & --- \\
 & Number of strategies & $H$ & 4 & --- \\
\bottomrule
\end{tabular}
\end{table}

\paragraph{Island growth model (Island):}\label{sec:island-model}

\begin{table}[!t]
\centering
\caption{Island model parameters for the pseudo-true data generation experiment.}
\label{tbl:island-params}
\begin{tabular}{@{}llccc@{}}
\toprule
Type & Parameter & Symbol & Pseudo-true value & Search bounds \\
\midrule
\multirow{7}{*}{Free (calibrated)} & Knowledge diffusion locality & $\rho$ & 0.01 & $[0,\,10]$ \\
 & Labour productivity & $\alpha$ & 1.5 & $[0.8,\,2]$ \\
 & Cumulative learning effect & $\varphi$ & 0.4 & $[0,\,1]$ \\
 & Island probability & $\pi$ & 0.4 & $[0,\,1]$ \\
 & Willingness to explore & $\varepsilon$ & 0.1 & $[0,\,1]$ \\
 & Initial number of firms & $m_0$ & 50 & $[10,\,100]$ \\
 & Poisson technology shock & $\lambda$ & 1.0 & $[0,\,1]$ \\
\midrule
\multirow{2}{*}{Simulation} & Series length & $T$ & 1000 & --- \\
\bottomrule
\end{tabular}
\end{table}

The Island growth model \cite{FagioloDosi2003,Lamperti2018} is an agent-based model of economic growth in which heterogeneous firms search for and exploit technological opportunities on a two-dimensional lattice. Each cell of the lattice contains an \emph{island} (a production opportunity with an associated productivity level) with probability $\pi$. A population of $m_0$ firms starts on a single home island. At each time step every firm follows one of three behaviours:

\begin{itemize}
  \item \textbf{Mining (exploitation):} the firm produces output on its current island, with returns to labour controlled by $\alpha$.
  \item \textbf{Exploring:} with probability $\varepsilon$, the firm leaves to search for a new unoccupied island; the productivity it unlocks scales with its accumulated experience through a learning effect $\varphi$.
  \item \textbf{Imitating:} after a transient period, the firm may move to a productive island it has heard about via a knowledge-diffusion signal whose locality is controlled by $\rho$.
\end{itemize}

\litrev{Each island can also receive a Poisson productivity shock with mean} $\lambda$\litrev{, representing an exogenous technological breakthrough. The observed output is aggregate Gross Domestic Product (GDP), defined as the total production across all active islands. It is simulated for} $T = 1000$ \litrev{periods. The seven free calibration parameters are} $\boldsymbol{\theta} = (\rho, \alpha, \varphi, \pi, \varepsilon, m_0, \lambda)$ \litrev{and are listed in} Table~\ref{tbl:island-params}. See Appendix~\ref{app:island} for the full formulation.

\subsection{Pseudo-true calibration protocol}\label{sec:expt-calibration}

\begin{algorithm}[!b]
\caption{Experiment setup per model (general)}
\label{alg:expt-setup}
\begin{algorithmic}[1]
\Require ABM model with parameter vector $\boldsymbol{\theta}$; calibration horizon $T$; objective set $\mathcal{O}$; optimizer configurations; number of independent runs $N_{\text{run}} \ge 1$.
\Ensure Summary of calibration performance (RMSE, timings) for all objective--optimizer--surrogate combinations.
\Statex
\State \textbf{Stage 0: Fixed settings}
\State Set number of simulation replications $R \gets 20$.
\State Define objective set $\mathcal{O} \gets \{\text{MSM}, \text{KS}, \text{MIC}, \text{GSL}\}$.
\State Define pure optimizers $\mathcal{A}_{\text{pure}} \gets \{\text{GA}, \text{PSO}\}$ and ML-assisted optimizers $\mathcal{A}_{\text{ML}} \gets \{\text{GA-ML}, \text{PSO-ML}\}$.
\State Define surrogate set $\mathcal{S} \gets \{\text{LR}, \text{SVM}, \text{RF}, \text{XGB}, \text{ANN}\}$.
\Statex
\State \textbf{Stage 1: Fix the pseudo-true parameters}
\State Choose ``true'' parameters $\boldsymbol{\theta}_{\text{true}}$ (default or target values); \rev{hold \mbox{$\boldsymbol{\theta}_{\text{true}}$} fixed across all runs.}
\Statex
\State \textbf{Stage 2: Calibration experiments}
\For{$run\_id = 1$ \textbf{to} $N_{\text{run}}$}
    \State \rev{Draw independent seeds \mbox{$(s^{\text{data}}_{run}, s^{\text{search}}_{run})$} from system entropy.}
    \State \rev{Simulate the ABM with \mbox{$\boldsymbol{\theta}_{\text{true}}$} for horizon \mbox{$T$} under seed \mbox{$s^{\text{data}}_{run}$} to obtain this run's pseudo-data \mbox{$\mathbf{y}^{(run)}_{\text{emp}}$}.}
    \State \rev{Derive the \mbox{$R$} replication seeds from \mbox{$s^{\text{search}}_{run}$}; share them across all configurations of this run.}
    \For{each configuration $(\ell, algo, surrogate)$ in $\mathcal{O} \times (\mathcal{A}_{\text{pure}} \cup \mathcal{A}_{\text{ML}}) \times \mathcal{S}$}
        \State Run calibration procedure \rev{\mbox{$\textsc{Calibrate}(\mathbf{y}^{(run)}_{\text{emp}}, T, \ell, algo, surrogate;\, s^{\text{search}}_{run})$}.}
        \State Obtain estimated parameters $\boldsymbol{\theta}_{\text{est}}$.
        \State Compute $\mathrm{RMSE} = \sqrt{\mathrm{mean}\bigl((\boldsymbol{\theta}_{\text{true}} - \boldsymbol{\theta}_{\text{est}})^2\bigr)}$.
        \State Record RMSE and wall-clock time for this configuration and run.
    \EndFor
\EndFor
\State Aggregate recorded results into a summary table or data set.
\State \Return summary
\end{algorithmic}
\end{algorithm}

\litrev{For each ABM, the protocol has three steps.} \rev{First, at the start of every independent run the ABM is simulated once with the fixed ``true'' parameters \mbox{$\boldsymbol{\theta}_{\text{true}}$} under that run's own pseudo-data seed, so each run calibrates against its own realisation \mbox{$\mathbf{y}^{(run)}_{\text{emp}}$} of the pseudo-ground-truth output.} \litrev{Second, each optimiser--surrogate--objective configuration is calibrated against this output to estimate} $\boldsymbol{\theta}_{\text{est}}$. \litrev{Third, parameter recovery accuracy is measured as} $\mathrm{RMSE}(\boldsymbol{\theta}_{\text{true}}, \boldsymbol{\theta}_{\text{est}})$.

\litrev{This pseudo-true design allows parameter recovery to be evaluated directly. Because} $\boldsymbol{\theta}_{\text{true}}$ \litrev{is known exactly by construction, the RMSE between} $\boldsymbol{\theta}_{\text{true}}$ \litrev{and the recovered estimate} $\boldsymbol{\theta}_{\text{est}}$ \litrev{provides a well-defined ground-truth measure of recovery accuracy. This measure is unavailable with real empirical data because the true parameters cannot be observed. The design therefore separates the effects of the optimiser, surrogate, and objective choices from the confounding effects of data uncertainty, enabling a direct comparison across all 48 configurations.} \rev{Because the target realisation is redrawn for every run while \mbox{$\boldsymbol{\theta}_{\text{true}}$} itself stays fixed, the repeated runs constitute a Monte-Carlo parameter-recovery experiment: the dispersion observed across runs measures the genuine sampling variability of the recovery exercise, not search noise alone.} Algorithm~\ref{alg:expt-setup} formalises the procedure.

\subsection{Experimental configuration}\label{sec:expt-config}

\begin{enumerate}
  \item \textbf{Test bed:} Two ABMs, BH (4 free parameters) and Island (7 free parameters), each evaluated on the 48 configurations. \rev{Each configuration is run \mbox{$N_{\text{run}} = 5$} times; every run redraws both the pseudo-data realisation and the search random-number stream from system entropy, while \mbox{$\boldsymbol{\theta}_{\text{true}}$}, all hyperparameters, the simulation budget \mbox{$N_{\text{ABM}}$}, and the top-50\% screening policy of \mbox{Section~\ref{sec:ml-algorithm}} are held fixed.}

  \item \textbf{Optimiser hyperparameters:} \litrev{The GA and PSO settings follow common defaults from the population-based search and SAEC literature. They are fixed across all runs and both ABMs.} Values are reported in Tables~\ref{tbl:ga-hp} and~\ref{tbl:pso-hp}.

  \item \textbf{Surrogate hyperparameters:} \litrev{All five surrogates (LR, SVM, XGBoost, RF, ANN) use their default settings. This keeps the comparison balanced and reproducible across all 48 configurations} (Table~\ref{tbl:ml-surrogates}).

  \item \textbf{Objective hyperparameters:} \litrev{The hyperparameters for each objective are fixed across both ABMs, allowing each objective to serve as a consistent comparator across all 48 configurations. The values are reported in} Table~\ref{tbl:objectives-hp}.

  \item \textbf{Replication and statistical protocol:} Each ABM evaluation averages over $R = 20$ Monte Carlo replications to control for ABM stochasticity, and each configuration is replicated across $N_{\text{run}} = 5$ independent calibration runs. \rev{Each run draws three independent sets of random seeds. The \emph{pseudo-data seed} generates the synthetic target series that the run calibrates against. The \emph{search seed} fixes the initial population or swarm, and every random choice the optimiser makes thereafter. From the search seed we derive the \mbox{$R = 20$} \emph{replication seeds} that drive the Monte Carlo replications of each ABM evaluation. Within a run, all 48 configurations share the same target series and the same replication seeds. They therefore solve an identical calibration problem and differ only in optimiser, surrogate and objective, so the run itself acts as a blocking factor. All three seeds are redrawn for the next run, which is what makes the five runs independent.} Statistical analysis in Section~\ref{sec:results-anova} uses one-way ANOVA with Dunnett's post-hoc tests at significance level $\alpha = 0.05$, with normality assessed via the Shapiro--Wilk test and homogeneity of variance via Levene's test. In all subsequent result tables the symbol $^{*}$ indicates $p < 0.05$.
\end{enumerate}

\begin{table}[H]
\begin{minipage}[t]{0.48\textwidth}
  \centering
  \captionof{table}{GA hyperparameters.}
  \label{tbl:ga-hp}
  \begin{tabular}{@{}lc@{}}
  \toprule
  Parameter & Value \\
  \midrule
  pop\_size & 50 \\
  max\_gen & 50 \\
  crossover\_rate & 0.7 \\
  mutation\_rate & 0.2 \\
  mutation\_std & 0.05 \\
  \bottomrule
  \end{tabular}
\end{minipage}\hfill
\begin{minipage}[t]{0.48\textwidth}
  \centering
  \captionof{table}{PSO hyperparameters.}
  \label{tbl:pso-hp}
  \begin{tabular}{@{}lc@{}}
  \toprule
  Parameter & Value \\
  \midrule
  swarm\_size & 50 \\
  max\_iters & 50 \\
  w\_inertia & 0.72 \\
  c1 & 1.49 \\
  c2 & 1.49 \\
  vmax\_factor & 0.25 \\
  \bottomrule
  \end{tabular}
\end{minipage}
\end{table}

\begin{table}[H]
  \centering
  \caption{Objective function hyperparameters.}
  \label{tbl:objectives-hp}
  \begin{tabular}{@{}llc@{}}
  \toprule
  Objective & Parameter & Value \\
  \midrule
  MSM & $L$ & $-1$ \\
  KS & $w_X$, $w_{|X|}$, $w_{X^2}$ & 1.0, 1.0, 1.0 \\
  MIC & $L, r$ & 3, 2 \\
  GSL-div & $L, b$ & 3, 4 \\
  \bottomrule
  \end{tabular}
\end{table}

\begin{table}[H]
\centering
\caption{ML surrogate hyperparameters.}
\label{tbl:ml-surrogates}
\small
\begin{tabular}{@{}llc@{\quad}llc@{}}
\toprule
Model & Parameter & Value & Parameter & Value \\
\midrule
XGBoost & eta & 0.3 & max\_depth & 6 \\
 & num\_rounds & 100 & subsample & 1 \\
 & colsample\_bytree & 1 & min\_child\_weight & 1 \\
 & lambda & 1 & alpha & 0 \\
 & seed & 1 & & \\
\midrule
Linear regression & learning\_rate & 0.1 & epochs & 50 \\
 & minibatch\_size & 1 & seed & 1 \\
\midrule
SVM & kernel\_type & RBF & seed & 1 \\
\midrule
Random forest & num\_trees & 100 & max\_depth & 10 \\
 & seed & 1 & fraction\_features\_in\_split & 0.5 \\
 & min\_child\_weight & 2 & min\_impurity\_decrease & 0 \\
\midrule
ANN & learning\_rate & 0.001 & hidden\_layer\_sizes & [64, 32] \\
 & num\_epochs & 100 & batch\_size & 32 \\
 & activation & RELU & updater & ADAM \\
 & seed & 1 & & \\
\bottomrule
\end{tabular}
\end{table}

\subsection{SHAP sensitivity analysis}\label{sec:expt-sensitivity}

\litrev{To complement the factorial calibration comparison, we conduct a post-experiment explainability analysis on the best-performing surrogate for each benchmark} (Section~\ref{sec:results}). The analysis uses SHAP (SHapley Additive exPlanations) \cite{Lundberg2017}. \litrev{SHAP assigns each ABM parameter a signed marginal contribution to the surrogate output. The sign shows whether the parameter, at its observed level, reduces or increases the calibration loss, while the magnitude measures the strength of the effect. Because SHAP values are computed directly from the trained surrogate, this sensitivity analysis adds negligible cost.}

\section{Results and discussion}\label{sec:results}

\litrev{This section evaluates the 48 calibration configurations defined in} Section~\ref{sec:experiments} \litrev{and formalised in} Algorithm~\ref{alg:expt-setup}. \litrev{Each configuration is assessed by two measures: parameter-recovery accuracy and wall-clock calibration time. Accuracy is measured as the RMSE between} $\boldsymbol{\theta}_{\text{true}}$ \litrev{and} $\boldsymbol{\theta}_{\text{est}}$\litrev{, which the pseudo-true design provides as a ground-truth criterion. We first present the Brock--Hommes results} (Section~\ref{sec:results-bh})\litrev{, followed by the Island growth model results} (Section~\ref{sec:results-island})\litrev{. We then use statistical tests to separate the surrogate effect from objective and run noise} (Section~\ref{sec:results-anova}). \litrev{Overall, inner-loop SAEC improves accuracy and speed simultaneously on both benchmarks. However, the best surrogate--optimiser--objective combination shifts predictably with ABM complexity.}

\subsection{Brock--Hommes model}\label{sec:results-bh}

\paragraph{Top-10 calibration outcomes:}\label{sec:results-bh-top10}
\litrev{The best configuration for the four-parameter BH model combines PSO-ML, the MIC objective, and an ANN surrogate. It achieves a mean RMSE of \mbox{$0.2286$} in \mbox{$5{,}932$}\,s. Compared with the strongest baseline without a surrogate (GA\,+\,MIC; RMSE \mbox{$0.2856$} in \mbox{$8{,}739$}\,s), this represents a \mbox{$20.0\%$} improvement in accuracy and a \mbox{$32.1\%$} improvement in speed. Overall, ML-assisted variants occupy eight of the top ten positions in \mbox{Table~\ref{tbl:bh-top10-rmse}}.}

\begin{figure}
  \centering
  \includegraphics[width=0.7\columnwidth]{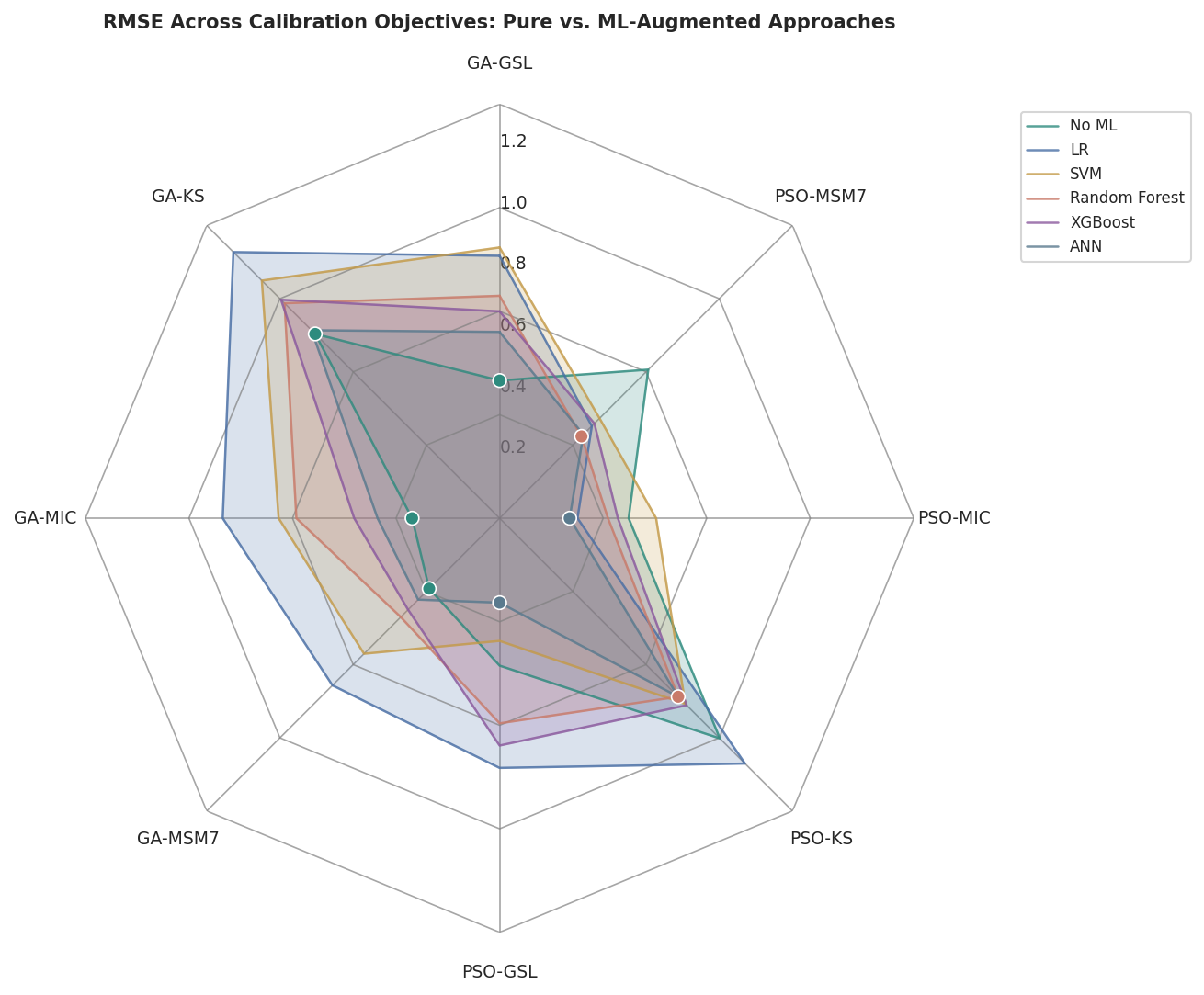}
  \caption{RMSE across calibration objectives for pure versus ML-augmented approaches on the BH model. Each axis represents an optimizer--objective pair; lower values (closer to the centre) indicate better parameter recovery. Circled points on each axis highlight the best-performing \rev{approach} for that optimizer--objective combination.}
  \label{fig:bh-radar-rmse}
\end{figure}

\litrev{However, the accuracy benefit depends on the optimiser. In \mbox{Figure~\ref{fig:bh-radar-rmse}}, a surrogate gives the best result on all four PSO axes: ANN on GSL-div (\mbox{$0.276$} versus \mbox{$0.481$} for pure PSO) and MIC (\mbox{$0.229$} versus \mbox{$0.421$}), and Random Forest on MSM and KS. The pattern reverses on all four GA axes, where the pure optimiser recovers the parameters more accurately than any surrogate-assisted variant. Thus, on BH, inner-loop screening improves accuracy only with PSO, although it reduces computation time with both optimisers. KS performs poorly throughout: its two axes have the largest radius of all objectives for both pure and surrogate-assisted approaches. The per-pair effects for every optimiser--objective combination are reported in \mbox{Appendix~\ref{app:tables}} (\mbox{Table~\ref{tbl:bh-improvement-accuracy}}).}

\litrev{The top ten results form two loose bands around the two pure-GA entries. Ranks \mbox{$1$}--\mbox{$3$} have RMSE values no greater than \mbox{$0.28$}. Ranks \mbox{$4$} and \mbox{$5$} have values of \mbox{$0.29$} and \mbox{$0.33$}, while ranks \mbox{$6$}--\mbox{$10$} range from \mbox{$0.35$} to \mbox{$0.39$}. PSO-ML holds the first three positions, suggesting that PSO's exploratory swarm dynamics work particularly well with an in-loop surrogate. Pure GA follows in fourth and fifth place with MIC and MSM, showing that a suitable objective can keep an unsurrogated optimiser competitive in a parameter space of this size. The second-ranked configuration is especially notable because it uses a \emph{linear} surrogate. A highly expressive model is therefore not necessary for effective screening, consistent with the later finding that surrogate fit and calibration quality can be weakly related.}

\litrev{These ranks should not be interpreted as strict performance differences. The standard deviations in \mbox{Table~\ref{tbl:bh-top10-rmse}} are similar in size to the gaps between configurations, and all ten overlap within one standard deviation. \mbox{Section~\ref{sec:results-anova}} tests which differences remain statistically distinguishable.}

\begin{table}[!htb]
\centering
\caption{Top-10 BH calibration configurations ranked by mean RMSE. \rev{Mean and standard deviation are taken over the five independent runs.}}
\label{tbl:bh-top10-rmse}
\begin{tabular}{@{}clllrrr@{}}
\toprule
Rank & Optimizer & Objective & ML Surrogate & Mean RMSE & \rev{SD} & Time (s) \\
\midrule
1\textsuperscript{$\star$} & \rev{PSO-ML} & \rev{MIC} & \rev{ANN} & \rev{\mbox{\textbf{0.2286}}} & \rev{0.1893} & \rev{5932} \\
2 & \rev{PSO-ML} & \rev{MIC} & \rev{LR} & \rev{0.2543} & \rev{0.1472} & \rev{5705} \\
3 & \rev{PSO-ML} & \rev{GSL-div} & \rev{ANN} & \rev{0.2758} & \rev{0.1667} & \rev{5940} \\
4 & \rev{GA} & \rev{MIC} & \rev{---} & \rev{0.2856} & \rev{0.2600} & \rev{8739} \\
5 & \rev{GA} & \rev{MSM} & \rev{---} & \rev{0.3251} & \rev{0.1567} & \rev{8734} \\
6 & \rev{PSO-ML} & \rev{MIC} & \rev{Random Forest} & \rev{0.3538} & \rev{0.2672} & \rev{5786} \\
7 & \rev{GA-ML} & \rev{MSM} & \rev{ANN} & \rev{0.3761} & \rev{0.1208} & \rev{\mbox{\textbf{5549}}} \\
8 & \rev{PSO-ML} & \rev{MSM} & \rev{Random Forest} & \rev{0.3780} & \rev{0.2563} & \rev{5807} \\
9 & \rev{PSO-ML} & \rev{MIC} & \rev{XGBoost} & \rev{0.3861} & \rev{0.3355} & \rev{5793} \\
10 & \rev{PSO-ML} & \rev{MSM} & \rev{ANN} & \rev{0.3864} & \rev{0.2114} & \rev{5860} \\
\bottomrule
\end{tabular}
\end{table}

\litrev{Computation time separates the two groups much more clearly than accuracy. Across the full factorial, ML-assisted configurations finish in \mbox{$5{,}277$}--\mbox{$5{,}940$}\,s, compared with \mbox{$8{,}734$}--\mbox{$9{,}215$}\,s for the pure optimisers. This is a mean saving of \mbox{$37.2\%$}, reaching \mbox{$42.7\%$} at the extremes. All forty ML-assisted configurations are faster than all eight pure baselines. The main reason is that the surrogate replaces roughly half of the ABM evaluations in each iteration under the top-\mbox{$50\%$} screening policy described in \mbox{Section~\ref{sec:ml-algorithm}}.}

\litrev{Within the ML-assisted group, GA-ML is slightly faster on average than PSO-ML (approximately \mbox{$5{,}474$}\,s versus \mbox{$5{,}789$}\,s). The fastest configuration overall is GA-ML\,+\,GSL-div\,+\,SVM at \mbox{$5{,}277$}\,s, although it does not appear in the accuracy top ten. Training costs differ substantially across surrogates: LR and SVM add about \mbox{$1$}--\mbox{$2$}\,s, XGBoost about \mbox{$29$}\,s, Random Forest \mbox{$202$}\,s, and ANN \mbox{$464$}\,s. These costs remain small relative to the ABM evaluations they replace. They therefore do not reorder the group: the fastest top-ten configuration is itself an ANN variant (rank~\mbox{$7$}, \mbox{$5{,}549$}\,s).}

\begin{figure}
  \centering
  \includegraphics[width=0.7\columnwidth]{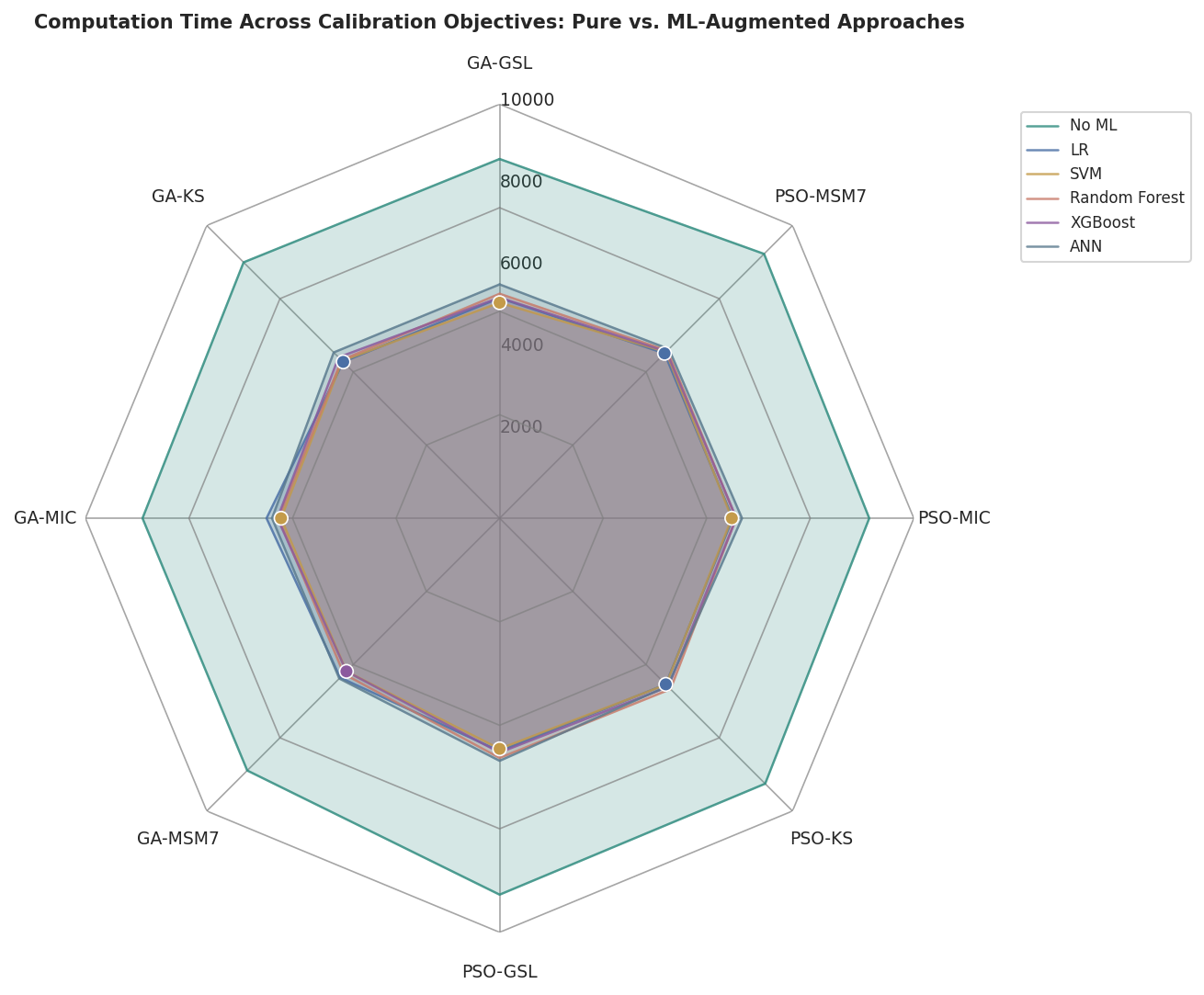}
  \caption{Computation time across calibration objectives for pure versus ML-augmented approaches on the BH model. Each axis represents an optimizer--objective pair; lower values (closer to the centre) indicate faster calibration. Circled points highlight the fastest surrogate for each optimizer--objective combination.}
  \label{fig:bh-radar-time}
\end{figure}

\litrev{The leading configuration offers the clearest practical result. PSO-ML\,+\,MIC\,+\,ANN is both \mbox{$32.1\%$} faster and \mbox{$20.0\%$} more accurate than the strongest pure baseline. It therefore improves accuracy and speed at the same time rather than trading one for the other. \mbox{Figure~\ref{fig:bh-radar-time}} shows the same pattern: ML-assisted approaches form a tighter, lower-cost envelope across all objectives. On seven of the eight axes, the fastest result comes from one of the two cheapest surrogates to train: SVM on four axes and LR on three. XGBoost is fastest on the remaining GA\,+\,MSM axis. Unlike the accuracy results in \mbox{Figure~\ref{fig:bh-radar-rmse}}, this timing pattern does not differ by optimiser.}

\begin{figure}[!htb]
  \centering
  \includegraphics[width=0.75\textwidth]{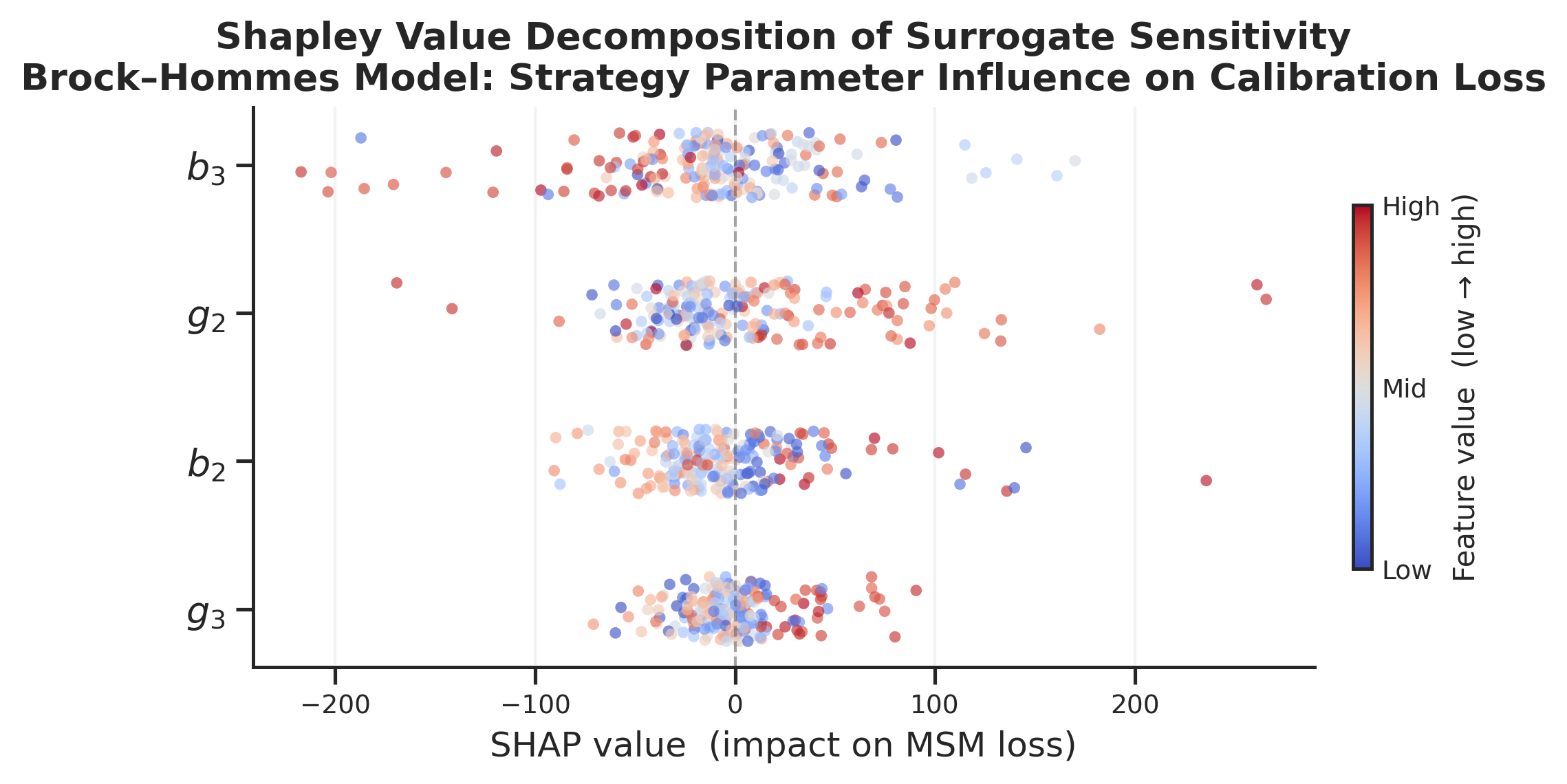}
  \caption{SHAP beeswarm plot for the surrogate of the \rev{single most accurate BH run (PSO-ML\,+\,MSM\,+\,XGBoost)}. Each dot is one \rev{of \mbox{$200$} parameter vectors sampled from the region the search covered}; horizontal position = signed Shapley value \rev{(negative reduces MSM loss, positive increases it)}; vertical axis ranks parameters by mean absolute influence\rev{, largest at the top}; colour encodes normalised parameter value (blue = low, red = high).}
  \label{fig:bh-shap-beeswarm}
\end{figure}

\paragraph{Parameter sensitivity:}\label{sec:results-bh-sensitivity}
\litrev{The SHAP analysis uses the surrogate from the single most accurate BH run: PSO-ML\,+\,MSM\,+\,XGBoost, with a parameter RMSE of \mbox{$0.0435$} (\mbox{Figure~\ref{fig:bh-shap-beeswarm}}). Influence is distributed across all four parameters rather than concentrated in one tier. The mean absolute Shapley values are \mbox{$37.4$} for \mbox{$b_3$}, \mbox{$35.9$} for \mbox{$g_2$}, \mbox{$27.8$} for \mbox{$b_2$}, and \mbox{$18.2$} for \mbox{$g_3$}. Thus, the most influential parameter has only about twice the weight of the least influential one.}

\litrev{The two bias terms account for \mbox{$54.7\%$} of total sensitivity, compared with \mbox{$45.3\%$} for the two trend coefficients. Because the ranking also interleaves the two parameter families, MSM constrains strategy switching and trend following to a similar degree. Of the four parameters, \mbox{$g_3$} is the least sharply identified.}

\litrev{The signed SHAP values show only a modest negative skew. All four parameters have negative medians, and \mbox{$57\%$}--\mbox{$64\%$} of the explained points lie below zero. Nevertheless, the distributions remain broadly symmetric around zero, and the mean for \mbox{$g_2$} is positive. The surrogate therefore identifies no single dominant direction of improvement across the search region. More precise identification of the trend coefficients would require an additional moment or a complementary objective.}

\begin{figure}[!htb]
  \centering
  \includegraphics[width=0.8\columnwidth]{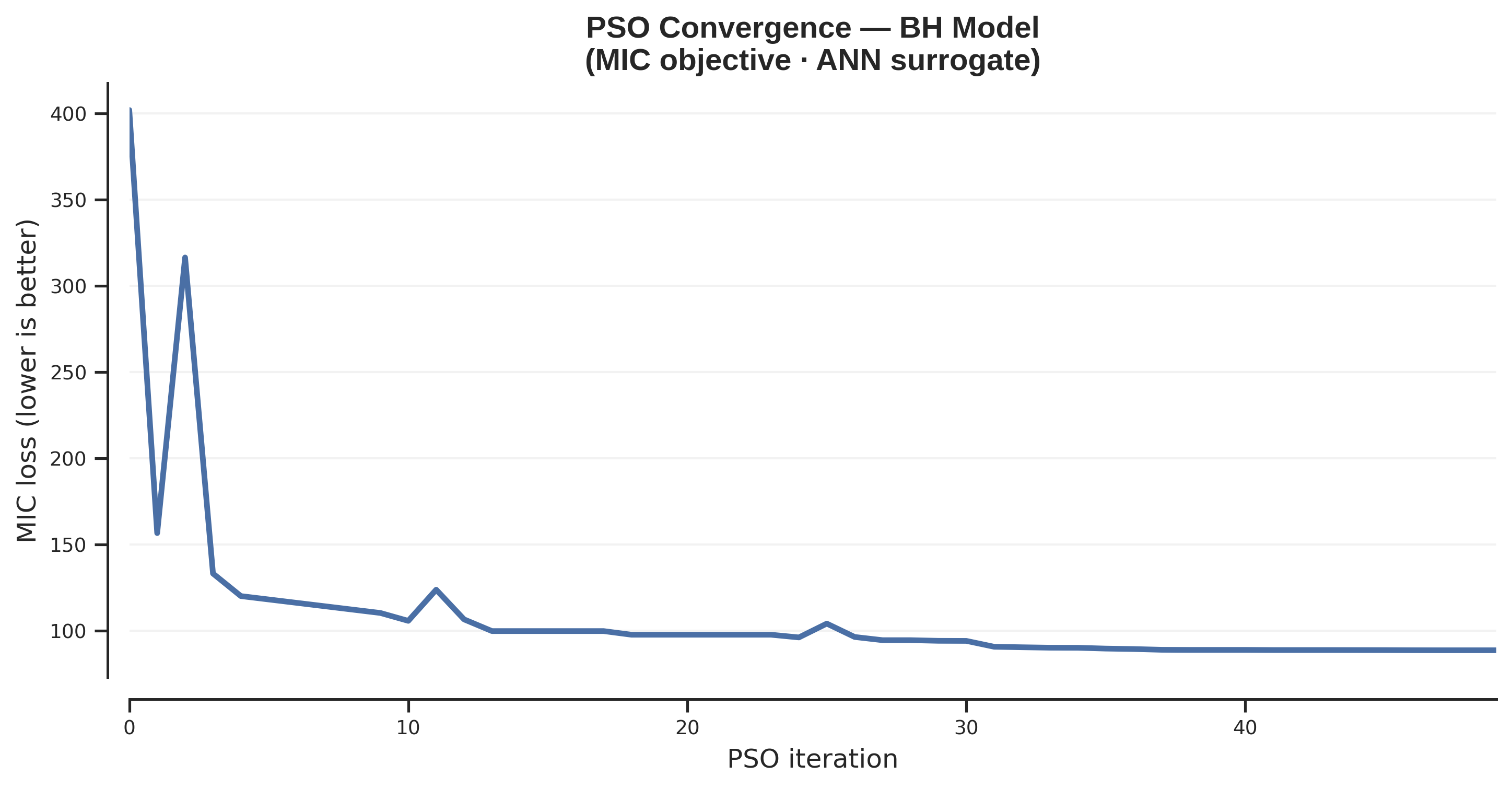}
  \caption{PSO convergence for the best BH configuration \rev{(PSO-ML\,+\,MIC\,+\,ANN)}. \rev{The curve traces the best-so-far MIC loss against PSO iteration; lower is better.}}
  \label{fig:bh-convergence-robustness}
\end{figure}

\paragraph{Robustness diagnostics:}\label{sec:results-bh-robustness}
\litrev{We assess the robustness of the main BH result in two ways: the convergence of the winning configuration and the predictive fit of the surrogates that guide the search. The best-so-far MIC loss for the rank-1 configuration falls sharply in the early iterations (\mbox{Figure~\ref{fig:bh-convergence-robustness}}). It reaches a stable plateau at about iteration~\mbox{$30$}, after which there is no material improvement.}

\litrev{Results still vary across runs. The five independent runs produce parameter RMSE values of \mbox{$0.09$}, \mbox{$0.11$}, \mbox{$0.35$}, \mbox{$0.50$}, and \mbox{$0.09$}, with a mean of \mbox{$0.229$} and a standard deviation of \mbox{$0.189$}. One run therefore settles in a visibly different basin. Because each run uses a different realisation of the pseudo-data, this spread reflects sampling variability in the recovery exercise rather than search noise. The run-level ANOVA in \mbox{Section~\ref{sec:results-anova}} uses this variability as its error term.}

\begin{figure}[!htb]
  \centering
  \includegraphics[width=0.8\columnwidth]{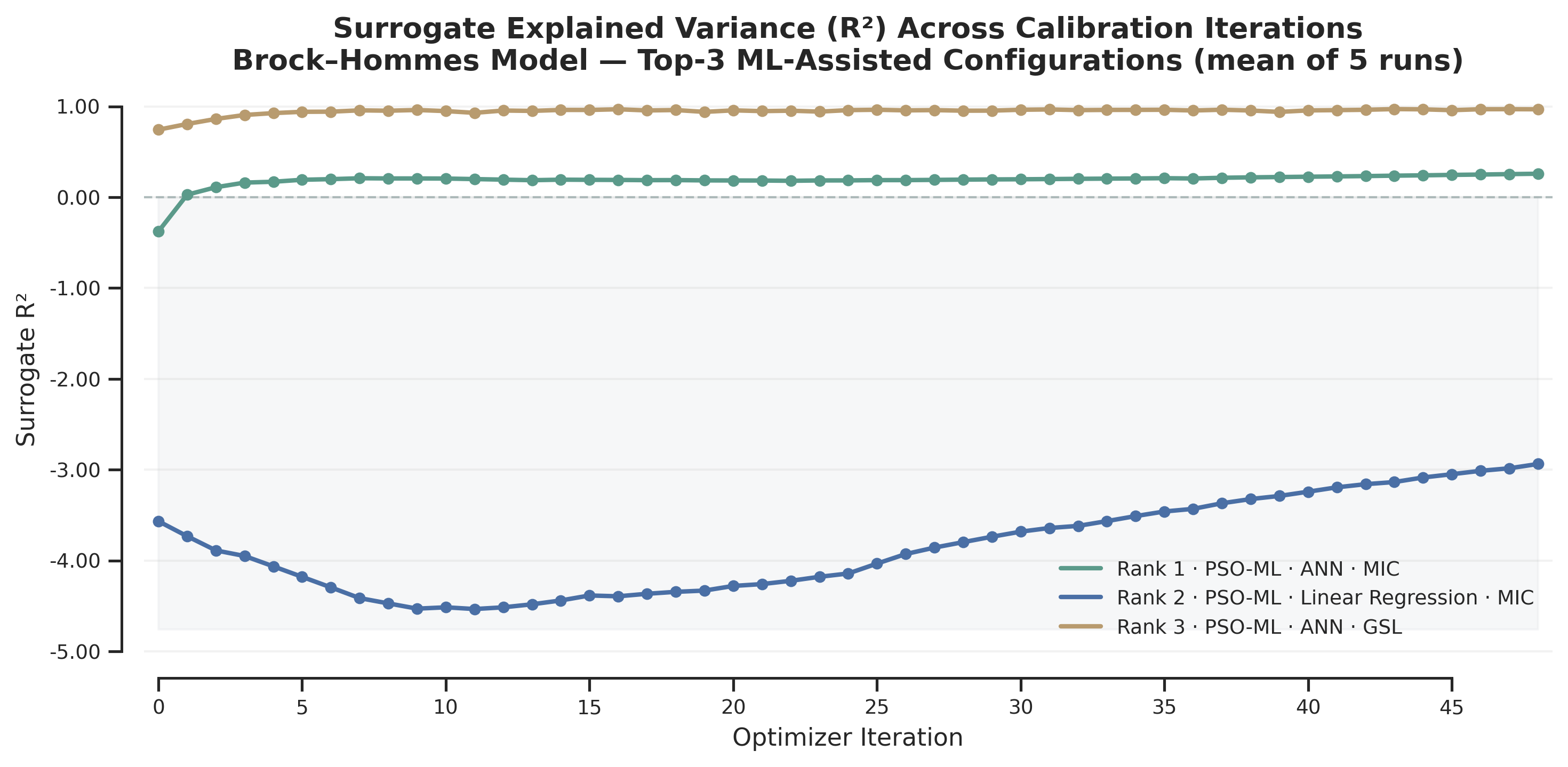}
  \caption{Surrogate $R^2$ across calibration iterations for the top-3 BH ML-assisted configurations, \rev{averaged over the five independent runs. Values below zero mark a surrogate whose squared error exceeds that of simply predicting the mean loss. Ranks 1 and 2 are fitted to MIC and rank 3 to GSL-div, so the three curves are not measured against a common loss surface.}}
  \label{fig:bh-surrogate-r2}
\end{figure}

\litrev{Surrogate fit tells a different story (\mbox{Figure~\ref{fig:bh-surrogate-r2}}), revealing a distinction between predictive accuracy and usefulness for calibration. Only the rank-3 ANN fitted to GSL-div behaves as a faithful emulator: its \mbox{$R^2$} rises from \mbox{$0.74$} to \mbox{$0.97$} and remains there. The rank-1 ANN on MIC levels off near \mbox{$0.20$}, so it explains only one fifth of the variation in the loss. The rank-2 linear surrogate has a negative \mbox{$R^2$} in all \mbox{$49$} iterations. It falls to \mbox{$-4.53$} by iteration~\mbox{$11$} and recovers to \mbox{$-2.94$}. A negative value means that its squared error is greater than that obtained by simply predicting the mean loss.}

\litrev{Ranks 1 and 2 provide a direct comparison because both use MIC. Replacing the linear surrogate with an ANN raises \mbox{$R^2$} from \mbox{$-2.94$} to \mbox{$0.26$}, but changes calibration RMSE only from \mbox{$0.2543$} to \mbox{$0.2286$}. Surrogate fit and calibration accuracy are therefore almost decoupled. This occurs because PSO uses the surrogate only to \emph{rank} candidates for ABM validation; it does not use the predicted loss itself. A surrogate can therefore guide the search successfully by ordering candidates reasonably well even when its loss predictions are poor. In this setting, goodness-of-fit is consequently a weak basis for choosing a surrogate, and the screening step can tolerate a much weaker emulator than a fit-based assessment would suggest.}

\subsection{Island growth model}\label{sec:results-island}

\paragraph{Top-10 calibration outcomes:}\label{sec:results-island-top10}
\litrev{For the seven-parameter Island model, the best configuration combines GA-ML, the KS objective, and an ANN surrogate. It achieves a mean RMSE of \mbox{$2.386$} in \mbox{$9{,}307$}\,s. Compared with the strongest pure baseline (GA\,+\,MSM; RMSE \mbox{$6.587$} in \mbox{$23{,}899$}\,s), this is a \mbox{$63.8\%$} improvement in accuracy and a \mbox{$61.1\%$} improvement in speed. ML-assisted variants occupy nine of the top ten positions in \mbox{Table~\ref{tbl:island-top10-rmse}}.}

\litrev{The optimiser pattern is the reverse of that observed on BH. PSO-ML holds the first three BH positions, whereas the GA family occupies eight of the ten Island positions, including first place. PSO-ML appears only at ranks \mbox{$6$} and \mbox{$7$}. This pattern suggests that GA performs more consistently across surrogate--objective combinations on the higher-dimensional Island landscape. \mbox{Appendix~\ref{app:tables}} reports the effect of each surrogate for every optimiser--objective combination (\mbox{Table~\ref{tbl:island-improvement-accuracy}}).}

\litrev{The Island results also show a clearer separation between objectives. KS occupies six of the top ten positions, including first place; GSL-div and MSM each occupy two. The two MSM entries reflect a broader exact tie: all six GA-family MSM configurations have an RMSE of \mbox{$6.587$} to four decimal places. Their order is therefore determined only by computation time, and only the two fastest appear in the top ten. No surrogate improves on pure GA with MSM. This result suggests that the MSM surface is sufficiently flat near the optimum for surrogate guidance to add nothing beyond the pure search. GSL-div and KS, by contrast, respond clearly to the choice of surrogate.}

\litrev{ANN and Random Forest are the most common surrogates in the top ten, with three entries each. LR appears twice and SVM once, while XGBoost, which led the pre-revision table, does not appear. The linear surrogate reaches rank~\mbox{$3$}, ahead of both Random Forest and SVM. As on BH, effective screening therefore does not require a highly expressive surrogate, which is consistent with the weak relationship between surrogate fit and calibration quality discussed below.}

\begin{figure}
  \centering
  \includegraphics[width=0.7\columnwidth]{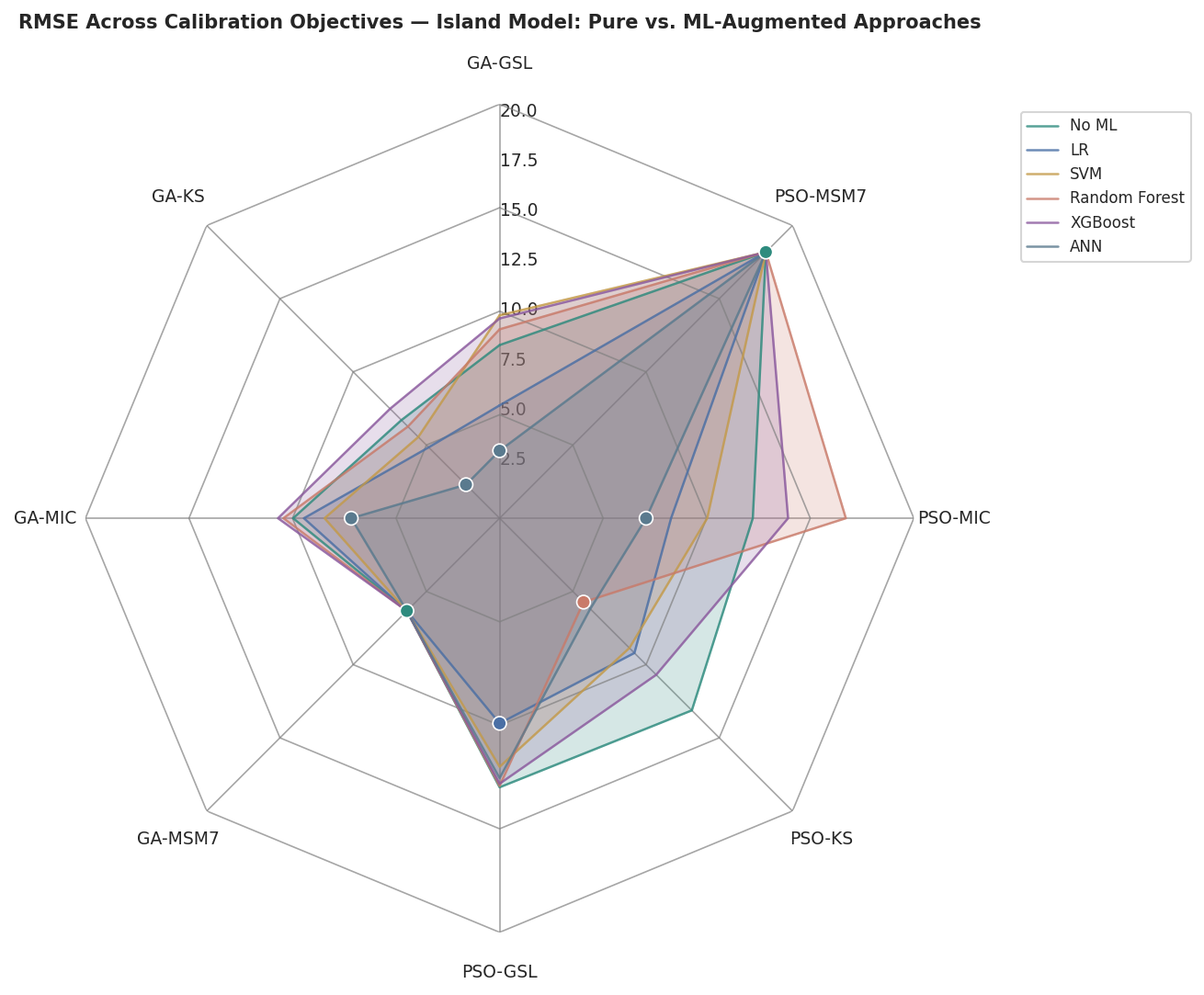}
  \caption{RMSE across calibration objectives for pure versus ML-augmented approaches on the Island model. Each axis represents an optimizer--objective pair; lower values (closer to the centre) indicate better parameter recovery. Circled points on each axis highlight the best-performing \rev{approach} for that optimizer--objective combination.}
  \label{fig:island-radar-rmse}
\end{figure}

\litrev{MIC is also absent from the top ten. In addition, all six PSO\,+\,MSM configurations have the same RMSE of \mbox{$18.912$}. This indicates a fundamental mismatch between PSO's search dynamics and the MSM objective on the Island landscape. \mbox{Figure~\ref{fig:island-radar-rmse}} confirms these patterns. The RMSE envelope moves inward on six of the eight axes, including both MIC axes: from \mbox{$10.371$} to \mbox{$7.452$} for GA and from \mbox{$12.716$} to \mbox{$7.357$} for PSO. Only the two MSM axes show no separation, with the pure and best-surrogate results coinciding exactly.}

\begin{table}[!htb]
\centering
\caption{Top-10 Island calibration configurations ranked by mean RMSE}
\label{tbl:island-top10-rmse}
\begin{tabular}{@{}clllrrr@{}}
\toprule
Rank & Optimizer & Objective & ML Surrogate & Mean RMSE & \rev{SD} & Time (s) \\
\midrule
1\textsuperscript{$\star$} & \rev{GA-ML} & \rev{KS} & \rev{ANN} & \rev{\mbox{\textbf{2.3862}}} & \rev{1.0717} & \rev{9307} \\
2 & \rev{GA-ML} & \rev{GSL-div} & \rev{ANN} & \rev{3.3922} & \rev{1.0623} & \rev{11615} \\
3 & \rev{GA-ML} & \rev{KS} & \rev{LR} & \rev{5.0387} & \rev{3.3567} & \rev{8935} \\
4 & \rev{GA-ML} & \rev{GSL-div} & \rev{LR} & \rev{5.6716} & \rev{2.4579} & \rev{16865} \\
5 & \rev{GA-ML} & \rev{KS} & \rev{SVM} & \rev{5.7752} & \rev{5.0327} & \rev{9558} \\
6 & \rev{PSO-ML} & \rev{KS} & \rev{Random Forest} & \rev{5.9595} & \rev{5.6590} & \rev{\mbox{\textbf{8645}}} \\
7 & \rev{PSO-ML} & \rev{KS} & \rev{ANN} & \rev{6.4392} & \rev{4.9591} & \rev{11555} \\
8 & \rev{GA-ML} & \rev{KS} & \rev{Random Forest} & \rev{6.5151} & \rev{3.8871} & \rev{11354} \\
9 & \rev{GA} & \rev{MSM} & \rev{---} & \rev{6.5874} & \rev{2.9427} & \rev{23899} \\
10 & \rev{GA-ML} & \rev{MSM} & \rev{Random Forest} & \rev{6.5874} & \rev{2.9427} & \rev{24331} \\
\bottomrule
\end{tabular}
\end{table}

\litrev{Computation times for the Island model form two distinct tiers. The eight KS and GSL-div configurations in the top ten finish in \mbox{$8{,}935$}--\mbox{$16{,}865$}\,s. The two MSM configurations require about \mbox{$24{,}000$}\,s. This gap is specific to GA. Averaged across surrogates, MSM is the most expensive GA objective at \mbox{$24{,}338$}\,s, compared with \mbox{$12{,}403$}\,s for KS, a ratio of \mbox{$2.0\times$}. Under PSO, however, MSM is the cheapest objective at \mbox{$8{,}490$}\,s, compared with \mbox{$19{,}979$}\,s for GSL-div.}

\litrev{For GA, nearly all of this difference comes from ABM simulation. MSM requires about \mbox{$24{,}326$}\,s of simulation time, whereas KS requires \mbox{$12{,}274$}\,s. MSM is therefore much more expensive on the Island model when paired with GA. This reverses the BH result, where MSM is among the fastest objectives and total time varies little by objective, ranging only from \mbox{$5{,}966$} to \mbox{$6{,}379$}\,s across the eight optimiser--objective pairs. A likely explanation is that the more rugged Island landscape generates more invalid evaluations and costly retries during MSM moment computation.}

\litrev{The fastest top-ten configuration is PSO-ML\,+\,KS\,+\,Random Forest (rank~\mbox{$6$}, \mbox{$8{,}645$}\,s). It is \mbox{$63.8\%$} faster than the only pure optimiser in the table. The most accurate configuration, GA-ML\,+\,KS\,+\,ANN (rank~\mbox{$1$}, \mbox{$9{,}307$}\,s), is simultaneously \mbox{$61.1\%$} faster and \mbox{$63.8\%$} more accurate than the best pure baseline. The joint improvement in accuracy and speed found on BH therefore also appears on Island, and is larger.}

\begin{figure}
  \centering
  \includegraphics[width=0.7\columnwidth]{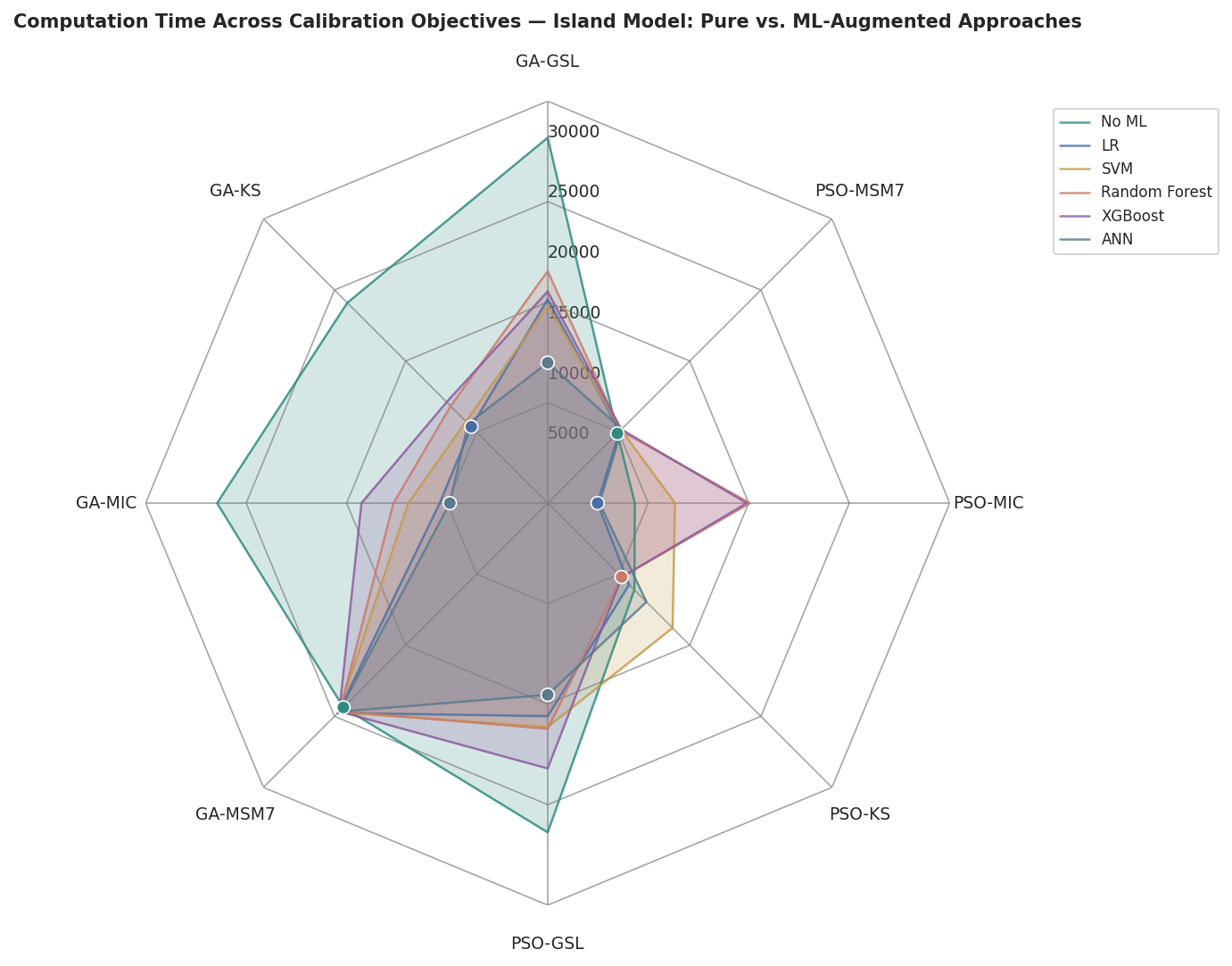}
  \caption{Computation time across calibration objectives for pure versus ML-augmented approaches on the Island model. Each axis represents an optimizer--objective pair; lower values (closer to the centre) indicate faster calibration. Circled points highlight the fastest \rev{approach} for each optimizer--objective combination.}
  \label{fig:island-radar-time}
\end{figure}

\litrev{On Island, objective choice affects wall-clock time more strongly than surrogate choice (\mbox{Figure~\ref{fig:island-radar-time}}). Relative to the fastest surrogate on each axis, the ML envelope moves inward on six of the eight axes. It loses only slightly on the two MSM axes (\mbox{$-1.8\%$} under GA and \mbox{$-3.5\%$} under PSO), which are almost flat across surrogates and match the exact RMSE ties reported above.}

\litrev{The size of the saving nevertheless depends strongly on the optimiser. The three non-MSM GA axes improve by \mbox{$62$}--\mbox{$70\%$}, compared with \mbox{$15$}--\mbox{$43\%$} for PSO. This difference arises because the pure PSO runs are already much faster: \mbox{$7{,}210$}--\mbox{$10{,}142$}\,s on MIC and KS, compared with \mbox{$23{,}441$}--\mbox{$30{,}233$}\,s for the corresponding pure GA runs. Surrogate choice also matters more than on BH. For PSO\,+\,MIC, the fastest surrogate finishes in \mbox{$4{,}127$}\,s and the slowest in \mbox{$16{,}735$}\,s, a fourfold difference. By comparison, the BH timings are relatively uniform: pure runs average about \mbox{$8{,}970$}\,s and ML-assisted runs about \mbox{$5{,}630$}\,s across all objectives.}

\begin{figure}[!htb]
  \centering
  \includegraphics[width=0.65\textwidth]{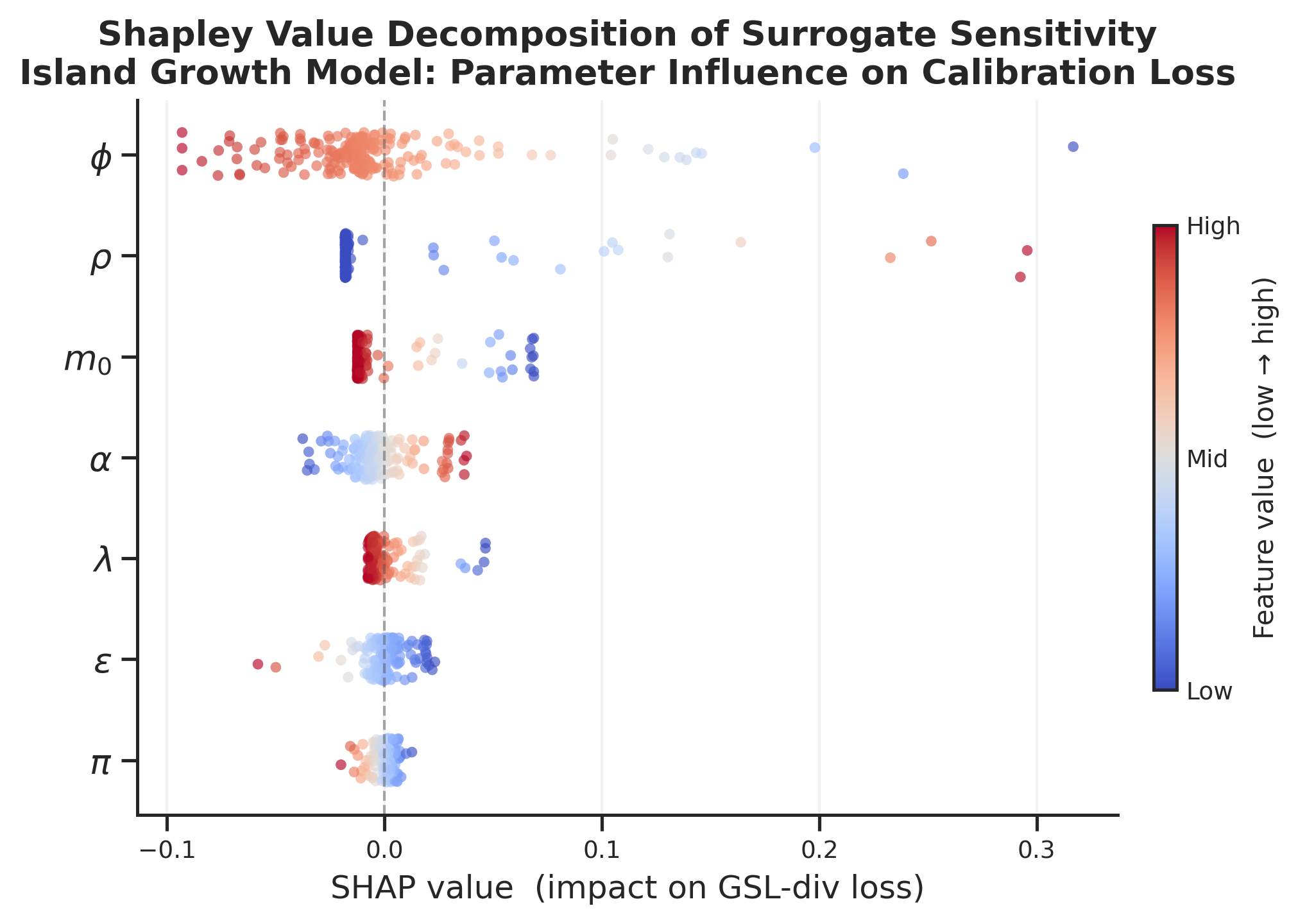}
  \caption{SHAP beeswarm plot for the surrogate of the \rev{single most accurate Island run (PSO-ML\,+\,GSL-div\,+\,LR)}. \rev{Each dot is one of \mbox{$200$} parameter vectors sampled from the region the search covered; horizontal position = signed Shapley value (negative reduces GSL-div loss, positive increases it); vertical axis ranks parameters by mean absolute influence, largest at the top; colour encodes normalised parameter value (blue = low, red = high).}}
  \label{fig:island-shap-beeswarm}
\end{figure}

\paragraph{Parameter sensitivity:}\label{sec:results-island-sensitivity}
\litrev{The SHAP analysis for the most accurate Island run identifies three levels of influence (\mbox{Figure~\ref{fig:island-shap-beeswarm}}). The cumulative learning parameter \mbox{$\varphi$} is most influential, accounting for \mbox{$30.9\%$} of the total. Knowledge-diffusion locality, \mbox{$\rho$}, follows at \mbox{$27.5\%$}. Together, these two parameters explain \mbox{$58.4\%$} of the total influence. This result reflects the model's design: \mbox{$\varphi$} governs how firms accumulate knowledge over time, while \mbox{$\rho$} represents the model's defining topology of localised, partly isolated knowledge pools.}

\litrev{The second tier contains the initial firm count \mbox{$m_0$} at \mbox{$16.0\%$} and returns to labour \mbox{$\alpha$} at \mbox{$10.4\%$}. The three most influential parameters together account for approximately \mbox{$74\%$}. The remaining parameters are technology-jump frequency \mbox{$\lambda$} (\mbox{$6.8\%$}), exploration willingness \mbox{$\varepsilon$} (\mbox{$5.3\%$}), and island probability \mbox{$\pi$} (\mbox{$3.1\%$}). Together, these three contribute \mbox{$15.2\%$}. They change the scale of the outcome without fundamentally reshaping the distributional pattern captured by GSL-div.}

\litrev{Because this surrogate is linear, the direction of each effect is exact rather than typical. Higher values of \mbox{$\varphi$}, \mbox{$\varepsilon$}, \mbox{$\lambda$}, \mbox{$m_0$}, and \mbox{$\pi$} reduce GSL-div loss. Higher values of \mbox{$\alpha$} and \mbox{$\rho$} increase it.}

\litrev{Two qualifications are important. First, influence is not the same as coefficient magnitude. Although \mbox{$\rho$} has a coefficient of only \mbox{$+0.033$}, it ranks second because the search covers a wide range, from \mbox{$0$} to \mbox{$9.37$}. Similarly, \mbox{$m_0$} ranks third despite a coefficient of \mbox{$-0.0009$}, because its search range extends from \mbox{$10.4$} to \mbox{$100$}. SHAP therefore measures sensitivity over the region visited by the search, not an intrinsic property of the model.}

\litrev{Second, the comparison with BH runs counter to what the higher dimensionality of Island might suggest. The four BH parameters have relatively even influence: the two bias terms account for \mbox{$54.7\%$} and the two trend parameters for \mbox{$45.3\%$}, while the most influential BH parameter has roughly twice the influence of the least. On Island, the most influential parameter has about ten times the influence of the least. Sensitivity is therefore more concentrated on Island, not less. The three least influential parameters remain weakly constrained by GSL-div alone and may require complementary objectives for precise identification.}

\begin{figure}[!htb]
  \centering
  \includegraphics[width=0.8\columnwidth]{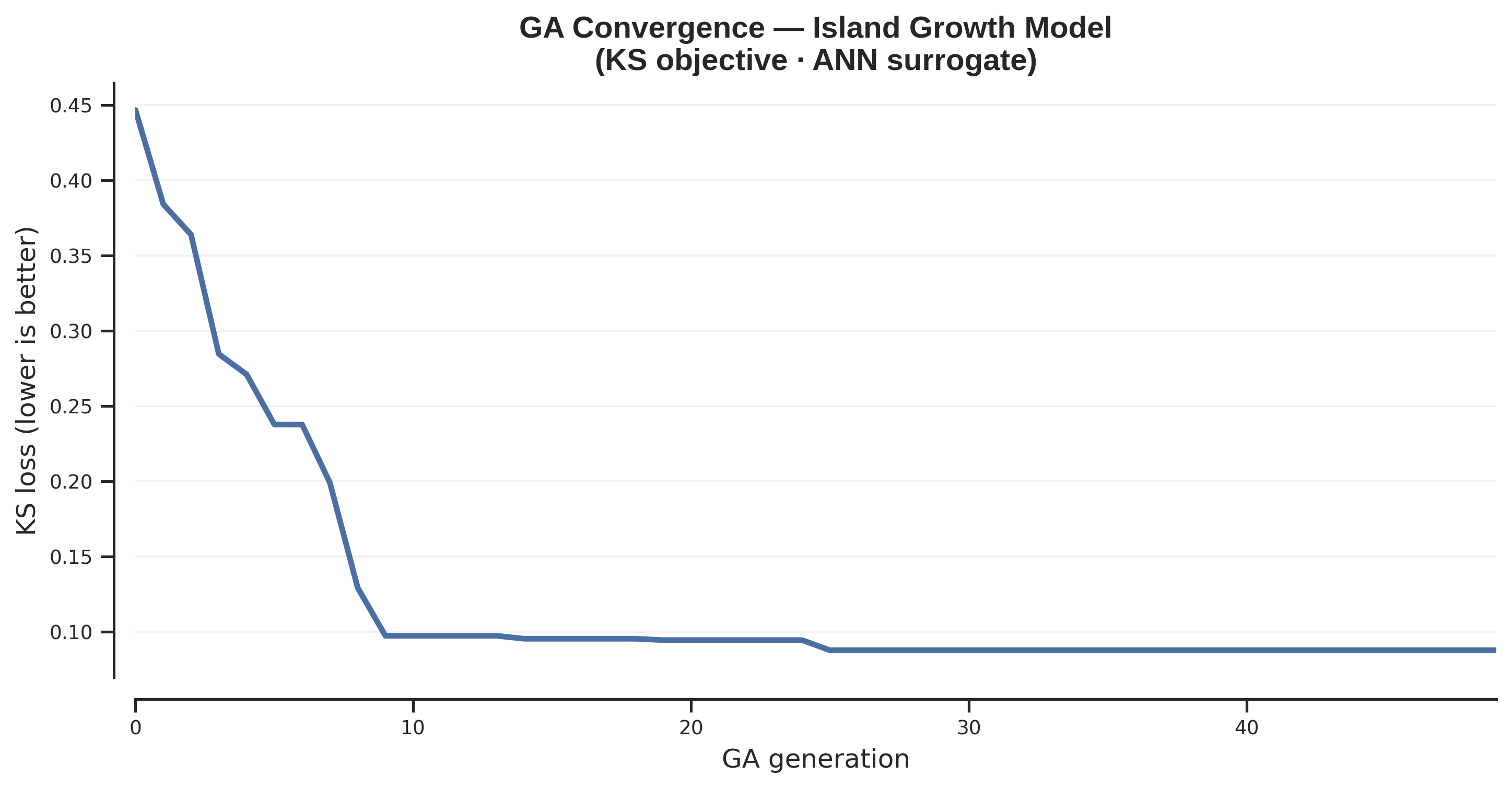}
  \caption{\rev{GA} convergence for the best Island configuration \rev{(GA-ML\,+\,KS\,+\,ANN)}. \rev{The curve traces the best-so-far KS loss against GA generation; lower is better.}}
  \label{fig:island-convergence-robustness}
\end{figure}

\paragraph{Robustness diagnostics:}\label{sec:results-island-robustness}
\litrev{We apply the same two robustness checks to the Island model. The best-so-far KS loss for the rank-1 configuration falls quickly at the start (\mbox{Figure~\ref{fig:island-convergence-robustness}}). It declines from \mbox{$0.447$} at generation~\mbox{$0$} to \mbox{$0.097$} by generation~\mbox{$9$}. Improvement is then small, with the final value of \mbox{$0.088$} reached at generation~\mbox{$25$}.}

\litrev{As on BH, the five runs do not converge to a single solution. The best configuration produces RMSE values of \mbox{$3.18$}, \mbox{$2.98$}, \mbox{$0.57$}, \mbox{$2.28$}, and \mbox{$2.93$}, with a mean of \mbox{$2.386$} and a standard deviation of \mbox{$1.072$}. This wider spread is consistent with the Island model's higher dimensionality and more rugged calibration landscape.}

\begin{figure}[!htb]
  \centering
  \includegraphics[width=0.8\columnwidth]{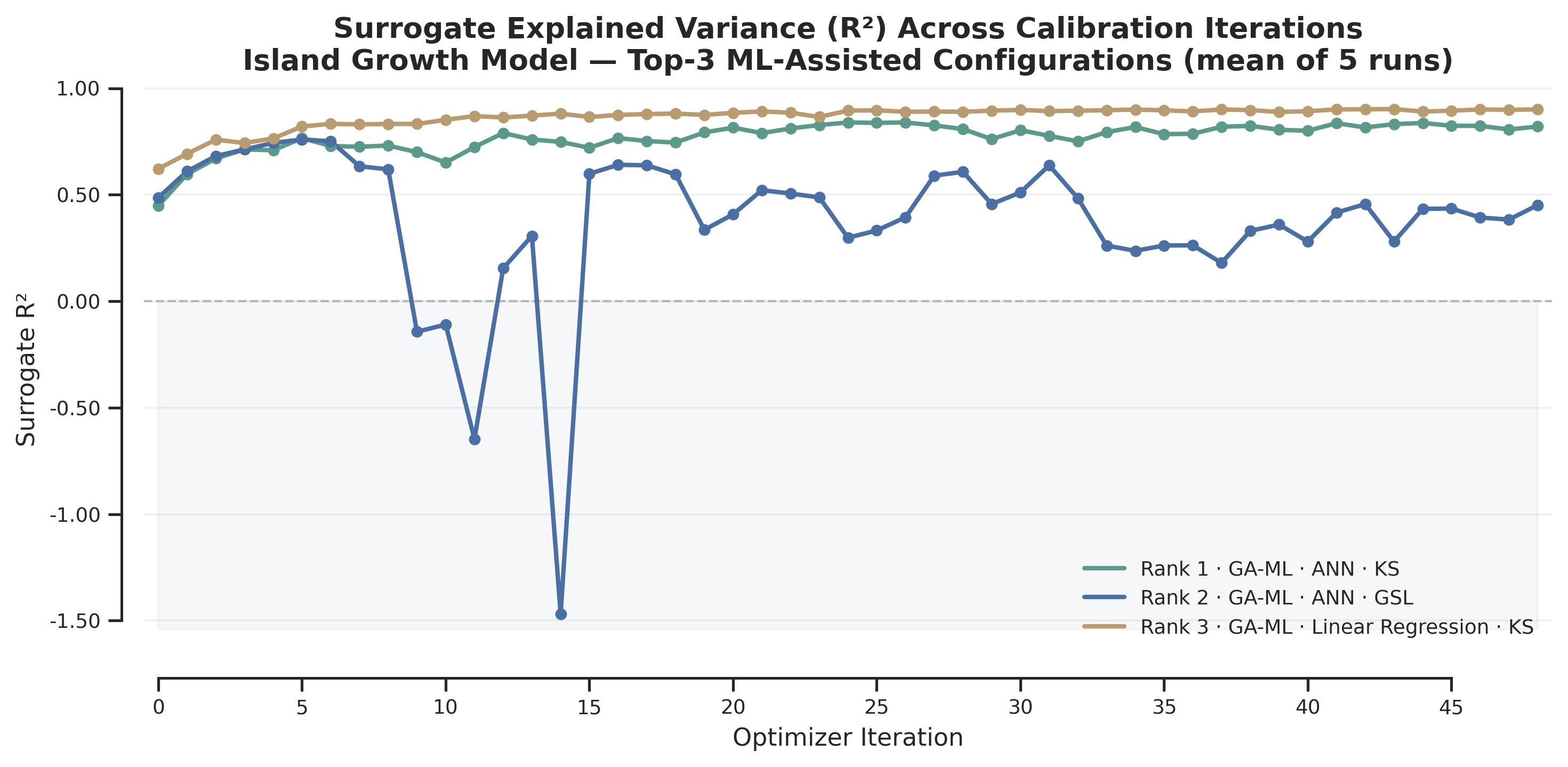}
  \caption{Surrogate $R^2$ across calibration iterations for the top-3 Island ML-assisted configurations\rev{, averaged over the five independent runs. Values below zero mark a surrogate whose squared error exceeds that of simply predicting the mean loss. Ranks 1 and 3 are fitted to KS and rank 2 to GSL-div, so the three curves are not measured against a common loss surface.}}
  \label{fig:island-surrogate-r2}
\end{figure}

\litrev{The predictive fit of the three leading surrogates also varies widely (\mbox{Figure~\ref{fig:island-surrogate-r2}}). Only the rank-3 linear surrogate behaves as a consistently faithful emulator, rising from \mbox{$R^2 = 0.62$} to \mbox{$0.90$} and remaining there. The rank-1 ANN improves steadily from \mbox{$0.45$} to \mbox{$0.82$}, but remains below \mbox{$0.80$} for \mbox{$27$} of the \mbox{$49$} generations. The rank-2 ANN, fitted to GSL-div, never exceeds \mbox{$0.76$}; its \mbox{$R^2$} becomes negative in four generations and reaches a minimum of \mbox{$-1.47$}.}

\litrev{Island therefore reproduces the BH pattern rather than reversing it. Surrogate fit and calibration accuracy have opposite rankings: the best-fitting of the three surrogates gives the worst parameter recovery, with an RMSE of \mbox{$5.04$}, compared with \mbox{$2.39$} for rank~\mbox{$1$}. As on BH, the search uses only the surrogate's ranking of candidates, not its predicted loss. Ranks 1 and 3 both use KS and can be compared directly. Rank 2 uses GSL-div, so its \mbox{$R^2$} is measured on a different loss surface.}

\subsection{Statistical confirmation}\label{sec:results-anova}

\litrev{The per-model results above identify the configurations that perform best on accuracy and speed. This subsection asks a narrower question: across objectives and independent runs, does adding a surrogate change parameter-recovery accuracy or computation time relative to the corresponding pure optimiser? We use the one-way ANOVA and Dunnett's post-hoc comparisons specified in \mbox{Section~\ref{sec:expt-config}} (item~\mbox{$5$}). Separate analyses are conducted for each model, optimiser, and outcome measure, with the pure optimiser as the control. Each group contains \mbox{$20$} observations from \mbox{$5$} independent runs across \mbox{$4$} objectives, giving six groups of \mbox{$n = 20$} in each analysis.}

\paragraph{Accuracy:}
\label{sec:results-anova-bh}\litrev{On BH, surrogate choice affects the two optimisers differently (\mbox{Table~\ref{tbl:bh-anova-rmse}}). For GA, the omnibus test rejects equality (\mbox{$F = 5.38$}, \mbox{$p < 0.001$}, \mbox{$\eta^2 = 0.19$}). LR (\mbox{$p < 0.001$}) and SVM (\mbox{$p = 0.004$}) significantly increase RMSE relative to pure GA, whereas Random Forest, XGBoost, and ANN do not differ significantly from the control. For PSO, the omnibus test does not reject equality (\mbox{$F = 1.17$}, \mbox{$p = 0.329$}, \mbox{$\eta^2 = 0.05$}), and no surrogate differs significantly from pure PSO. Four surrogates reduce mean RMSE; ANN gives the largest reduction (\mbox{$-0.223$}), but it is not significant (\mbox{$p = 0.169$}). LR is effectively identical to the PSO control (\mbox{$+0.006$}, \mbox{$p = 1.000$}). Thus, the adverse accuracy effects on BH are confined to LR and SVM under GA.}

\begin{table}[!htb]
  \centering
  \caption{Dunnett's post-hoc test results for calibration RMSE on the Brock--Hommes model. The control mean for GA is \rev{$0.478$} and for PSO is \rev{$0.651$}. $\Delta$ denotes the difference (treatment $-$ control); $^{*}$ indicates $p < 0.05$. \rev{Positive \mbox{$\Delta$} means the surrogate is less accurate than the pure optimiser.}}
  \label{tbl:bh-anova-rmse}
  \begin{tabular}{@{}lrrrlrrl@{}}
    \toprule
    & \multicolumn{3}{c}{GA} & & \multicolumn{3}{c}{PSO} \\
    \cmidrule(lr){2-4} \cmidrule(l){6-8}
    Surrogate & Mean & $\Delta$ & $p$ & & Mean & $\Delta$ & $p$ \\
    \midrule
    LR           & \rev{0.940} & \rev{\mbox{$+$0.462}} & \rev{\mbox{${<}0.001^{*}$}} & & \rev{0.657} & \rev{\mbox{$+$0.006}} & \rev{1.000} \\
    SVM          & \rev{0.832} & \rev{\mbox{$+$0.354}} & \rev{\mbox{$0.004^{*}$}}    & & \rev{0.558} & \rev{\mbox{$-$0.094}} & \rev{0.864} \\
    RandomForest & \rev{0.710} & \rev{\mbox{$+$0.232}} & \rev{0.108}                 & & \rev{0.556} & \rev{\mbox{$-$0.095}} & \rev{0.859} \\
    XGBoost      & \rev{0.645} & \rev{\mbox{$+$0.168}} & \rev{0.359}                 & & \rev{0.607} & \rev{\mbox{$-$0.044}} & \rev{0.994} \\
    ANN          & \rev{0.563} & \rev{\mbox{$+$0.085}} & \rev{0.883}                 & & \rev{0.429} & \rev{\mbox{$-$0.223}} & \rev{0.169} \\
    \bottomrule
  \end{tabular}
\end{table}

\label{sec:results-anova-island}\litrev{On Island, neither omnibus test rejects equality (GA: \mbox{$F = 2.20$}, \mbox{$p = 0.060$}; PSO: \mbox{$F = 1.24$}, \mbox{$p = 0.294$}), and no surrogate differs significantly from either control (\mbox{Table~\ref{tbl:island-anova-rmse}}). The mean effects nevertheless tend to favour surrogate assistance: all five surrogates reduce mean RMSE under PSO, and three do so under GA. ANN gives the largest reduction for both optimisers: \mbox{$39.3\%$} under GA (\mbox{$\Delta=-3.20$}, \mbox{$p = 0.081$}) and \mbox{$22.2\%$} under PSO (\mbox{$\Delta=-3.26$}, \mbox{$p = 0.228$}), but neither comparison reaches significance. Group standard deviations of \mbox{$3.7$}--\mbox{$6.2$} RMSE units exceed the differences between group means (\mbox{$0.2$}--\mbox{$3.3$}), so the observed mean gains cannot be separated from run-to-run variation at this sample size.}

\begin{table}[!htb]
  \centering
  \caption{Dunnett's post-hoc test results for calibration RMSE on the Island growth model. The control mean for GA is \rev{$8.158$} and for PSO is \rev{$14.700$}. $\Delta$ denotes the difference (treatment $-$ control); negative values indicate better parameter recovery than the pure optimiser. \rev{No comparison reaches \mbox{$p < 0.05$} under either optimiser, so no entry is starred.}}
  \label{tbl:island-anova-rmse}
  \begin{tabular}{@{}lrrrlrrl@{}}
    \toprule
    & \multicolumn{3}{c}{GA} & & \multicolumn{3}{c}{PSO} \\
    \cmidrule(lr){2-4} \cmidrule(l){6-8}
    Surrogate & Mean & $\Delta$ & $p$ & & Mean & $\Delta$ & $p$ \\
    \midrule
    LR           & \rev{6.782}  & \rev{\mbox{$-$1.376}} & \rev{0.768} & & \rev{11.851} & \rev{\mbox{$-$2.849}} & \rev{0.346} \\
    SVM          & \rev{7.838}  & \rev{\mbox{$-$0.320}} & \rev{0.999} & & \rev{12.762} & \rev{\mbox{$-$1.938}} & \rev{0.699} \\
    RandomForest & \rev{8.363}  & \rev{\mbox{$+$0.205}} & \rev{1.000} & & \rev{13.928} & \rev{\mbox{$-$0.772}} & \rev{0.990} \\
    XGBoost      & \rev{8.884}  & \rev{\mbox{$+$0.726}} & \rev{0.978} & & \rev{14.465} & \rev{\mbox{$-$0.235}} & \rev{1.000} \\
    ANN          & \rev{4.955}  & \rev{\mbox{$-$3.204}} & \rev{0.081} & & \rev{11.442} & \rev{\mbox{$-$3.258}} & \rev{0.228} \\
    \bottomrule
  \end{tabular}
\end{table}

\litrev{Taken together, the accuracy tests identify only two adverse effects: LR and SVM under GA on BH, and no significant beneficial effects. The evidence therefore supports the narrower conclusion that inner-loop screening generally preserves parameter-recovery accuracy, not the stronger claim that it systematically improves it.}

\paragraph{Computation time:}
\litrev{The BH timing results are much more uniform (\mbox{Table~\ref{tbl:bh-anova-time}}). Every surrogate reduces computation time significantly relative to both pure GA and pure PSO (\mbox{$p < 0.001$} in all cases). The omnibus statistics are \mbox{$F = 1402$} for GA and \mbox{$F = 2344$} for PSO, with \mbox{$\eta^2 = 0.98$} and \mbox{$0.99$}, respectively. Savings range from \mbox{$35.7$}--\mbox{$39.1\%$} for GA and \mbox{$35.7$}--\mbox{$37.6\%$} for PSO. Even ANN, the most costly surrogate to train, saves \mbox{$35.7\%$} under both optimisers: its training overhead of about \mbox{$460$}\,s remains small relative to the \mbox{$3{,}100$}--\mbox{$3{,}400$}\,s of ABM evaluation it replaces.}

\begin{table}[!htb]
  \centering
  \caption{Dunnett's post-hoc test results for total computation time (seconds) on the Brock--Hommes model. The control mean for GA is \rev{$8{,}782$}\,s and for PSO is \rev{$9{,}151$}\,s. $\Delta$ denotes time saved (negative values indicate faster runtime); all $p < 0.001^{*}$.}
  \label{tbl:bh-anova-time}
  \begin{tabular}{@{}lrrlrrl@{}}
    \toprule
    & \multicolumn{3}{c}{GA} & \multicolumn{3}{c}{PSO} \\
    \cmidrule(lr){2-4} \cmidrule(l){5-7}
    Surrogate & Mean (s) & $\Delta$ (s) & $p$ & Mean (s) & $\Delta$ (s) & $p$ \\
    \midrule
    LR           & \rev{5{,}507} & \rev{\mbox{$-$3{,}275}} & ${<}0.001^{*}$ & \rev{5{,}717} & \rev{\mbox{$-$3{,}434}} & ${<}0.001^{*}$ \\
    SVM          & \rev{5{,}345} & \rev{\mbox{$-$3{,}438}} & ${<}0.001^{*}$ & \rev{5{,}712} & \rev{\mbox{$-$3{,}439}} & ${<}0.001^{*}$ \\
    RandomForest & \rev{5{,}444} & \rev{\mbox{$-$3{,}338}} & ${<}0.001^{*}$ & \rev{5{,}844} & \rev{\mbox{$-$3{,}307}} & ${<}0.001^{*}$ \\
    XGBoost      & \rev{5{,}429} & \rev{\mbox{$-$3{,}353}} & ${<}0.001^{*}$ & \rev{5{,}785} & \rev{\mbox{$-$3{,}366}} & ${<}0.001^{*}$ \\
    ANN          & \rev{5{,}646} & \rev{\mbox{$-$3{,}136}} & ${<}0.001^{*}$ & \rev{5{,}887} & \rev{\mbox{$-$3{,}264}} & ${<}0.001^{*}$ \\
    \bottomrule
  \end{tabular}
\end{table}

\litrev{On Island, the timing effect depends on the optimiser (\mbox{Table~\ref{tbl:island-anova-time}}). Under GA, the omnibus test rejects equality (\mbox{$F = 6.80$}, \mbox{$p < 0.001$}, \mbox{$\eta^2 = 0.23$}), and all five surrogates reduce time significantly (\mbox{$p \le 0.002$}). ANN gives the largest saving (\mbox{$49.1\%$}), followed by LR (\mbox{$43.5\%$}), SVM (\mbox{$41.0\%$}), Random Forest (\mbox{$35.5\%$}), and XGBoost (\mbox{$34.1\%$}). Under PSO, neither the omnibus test (\mbox{$F = 0.72$}, \mbox{$p = 0.611$}, \mbox{$\eta^2 = 0.03$}) nor any post-hoc comparison is significant. LR and ANN provide mean savings of \mbox{$24.7\%$} and \mbox{$23.5\%$}; SVM and Random Forest save less than \mbox{$1\%$}, whereas XGBoost is \mbox{$741$}\,s slower. Pure PSO is already a strong baseline at \mbox{$13{,}185$}\,s, compared with \mbox{$26{,}230$}\,s for pure GA.}

\begin{table}[!htb]
  \centering
  \caption{Dunnett's post-hoc test results for total computation time (seconds) on the Island growth model. The control mean for GA is \rev{$26{,}230$}\,s and for PSO is \rev{$13{,}185$}\,s. $\Delta$ denotes time saved (negative values indicate faster runtime); $^{*}$ indicates $p < 0.05$. \rev{All five GA comparisons are significant; none of the PSO comparisons is.}}
  \label{tbl:island-anova-time}
  \begin{tabular}{@{}lrrlrrl@{}}
    \toprule
    & \multicolumn{3}{c}{GA} & \multicolumn{3}{c}{PSO} \\
    \cmidrule(lr){2-4} \cmidrule(l){5-7}
    Surrogate & Mean (s) & $\Delta$ (s) & $p$ & Mean (s) & $\Delta$ (s) & $p$ \\
    \midrule
    LR           & \rev{14{,}817} & \rev{\mbox{$-$11{,}413}} & \rev{\mbox{${<}0.001^{*}$}} & \rev{9{,}930}  & \rev{\mbox{$-$3{,}255}} & \rev{0.691} \\
    SVM          & \rev{15{,}475} & \rev{\mbox{$-$10{,}755}} & \rev{\mbox{${<}0.001^{*}$}} & \rev{13{,}061} & \rev{\mbox{$-$123}}     & \rev{1.000} \\
    RandomForest & \rev{16{,}909} & \rev{\mbox{$-$9{,}322}}  & \rev{\mbox{$0.001^{*}$}}    & \rev{13{,}139} & \rev{\mbox{$-$46}}      & \rev{1.000} \\
    XGBoost      & \rev{17{,}280} & \rev{\mbox{$-$8{,}950}}  & \rev{\mbox{$0.002^{*}$}}    & \rev{13{,}926} & \rev{\mbox{$+$741}}     & \rev{0.999} \\
    ANN          & \rev{13{,}355} & \rev{\mbox{$-$12{,}875}} & \rev{\mbox{${<}0.001^{*}$}} & \rev{10{,}090} & \rev{\mbox{$-$3{,}095}} & \rev{0.730} \\
    \bottomrule
  \end{tabular}
\end{table}

\litrev{The Island PSO result reflects variation in evaluation cost rather than call count alone. Screening reduces ABM evaluations from \mbox{$2{,}550$} to about \mbox{$1{,}375$} (\mbox{$46\%$}), but \mbox{$m_0$} controls the initial number of firms simulated for \mbox{$T=1000$} periods across \mbox{$20$} replications. The evaluated candidates shift from a mean \mbox{$m_0$} of about \mbox{$48$} under pure PSO to about \mbox{$68$} under PSO-ML, so screening selects fewer but more expensive simulations. Under GA, the search remains near \mbox{$m_0=66$}, allowing the reduction in calls to produce the observed \mbox{$34$}--\mbox{$49\%$} saving. BH has no parameter that scales simulation cost and consequently shows more uniform \mbox{$35$}--\mbox{$39\%$} savings.}

\litrev{Taken together, the tests confirm a computation-time benefit in three of the four model--optimiser analyses and no significant difference in the fourth. They provide no evidence that surrogates systematically improve parameter-recovery accuracy: apart from the adverse LR and SVM effects under GA on BH, accuracy is statistically indistinguishable from the pure controls. The defensible conclusion is therefore that inner-loop surrogate screening reduces computation time while generally preserving accuracy. The large accuracy gains in \mbox{Sections~\ref{sec:results-bh}} and \mbox{\ref{sec:results-island}} remain best-configuration outcomes rather than average effects separated from run-to-run variation.}

\section{Conclusion}\label{sec:conclusion}

\litrev{Agent-based model calibration is often constrained by objective evaluations that are both computationally expensive and stochastic. This study adapts \emph{inner-loop SAEC} to this setting by embedding a machine-learning surrogate within genetic-algorithm and particle-swarm search. At each iteration, the surrogate screens the candidates and the simulator validates only the top \mbox{$50\%$} under a fixed simulation budget. The evaluation covers \mbox{$48$} configurations: \mbox{$8$} pure-optimiser baselines and \mbox{$40$} ML-assisted variants, across two optimisers, five surrogates, and four objectives on the Brock--Hommes asset-pricing and Island growth models.}

\litrev{The best configuration improves accuracy and speed simultaneously on both benchmarks. PSO-ML\,+\,MIC\,+\,ANN reduces parameter-recovery RMSE by \mbox{$20.0\%$} and calibration time by \mbox{$32.1\%$} on Brock--Hommes, while GA-ML\,+\,KS\,+\,ANN reduces them by \mbox{$63.8\%$} and \mbox{$61.1\%$}, respectively, on Island. These results avoid an accuracy--speed trade-off, but they do not identify a universal recipe: PSO leads on the four-parameter Brock--Hommes landscape, whereas GA leads on the seven-parameter Island landscape. Performance therefore depends on the interaction among the model, optimiser, objective, and surrogate.}

\litrev{The statistical analysis provides strong evidence for the computational benefit. Every surrogate saves significant time under both Brock--Hommes optimisers and under GA on Island, where ANN gives the largest saving at \mbox{$49.1\%$}. The accuracy evidence is more cautious: no surrogate has a significant effect on Island parameter recovery, and only LR and SVM have significant adverse effects under GA on Brock--Hommes. The headline accuracy gains should therefore be read as best-configuration comparisons rather than average effects separated from run-to-run variation. The SHAP analysis adds parameter-identifiability information without further simulation: two of four Brock--Hommes parameters account for \mbox{$61.4\%$} of sensitivity, while three of seven Island parameters account for approximately \mbox{$74\%$}.}

\litrev{Three conditions define the framework's \emph{operating envelope}. Default surrogate hyperparameters keep the factorial balanced and reproducible, making the reported gains conservative with respect to tuning. The fixed \mbox{$50\%$} screening fraction provides a clear control but leaves adaptive policies untested. Finally, the advantage depends on ABM evaluations remaining more expensive than surrogate training and prediction; it narrows when the objective is inexpensive or surrogate overhead approaches simulation cost, which also limits the practical choice of surrogate architecture.}

\litrev{Future work should therefore examine targeted surrogate tuning and adaptive top-\mbox{$k$} policies based on confidence or search progress. Broader ABM classes, multi-objective and uncertainty-aware calibration, and online selection from surrogate ensembles would further test the framework where one fixed surrogate--objective pair is insufficient, while seeking to preserve the cost and accuracy benefits demonstrated here.}

\section*{CRediT authorship contribution statement}

\textbf{Duguma Yeshitla Habtemariam:} Conceptualization, Methodology, Experimentation, Analysis, Writing. \textbf{Jihwan Lee:} Supervision, Writing, Proofreading.

\section*{Declaration of generative AI and AI-assisted technologies in the manuscript preparation process}

During the preparation of this work the authors used Claude AI in order to improve the readability and conciseness of the manuscript. After using this tool, the authors reviewed and edited the content as needed and take full responsibility for the content of the published article.

\appendix
\counterwithin{figure}{section}
\counterwithin{table}{section}
\renewcommand{\thefigure}{\thesection\arabic{figure}}
\renewcommand{\thetable}{\thesection\arabic{table}}
\section*{Appendix}

\section{Definitions}\label{app:definitions}

\subsection{Method of simulated moments (MSM)}\label{app:msm}

The method of simulated moments (MSM) is an SMD-style objective that originates from the method of moments (MM) \cite{Alahmadi2024}. The basic idea is to select a set of informative \emph{moment conditions} (summary statistics) from the data and then choose parameters so that the same moments computed from model simulations match the empirical moments as closely as possible \cite{GrazziniRichiardi2015}. In its generic quadratic-form version, letting $\mathbf{m}_D \in \mathbb{R}^{p}$ denote the vector of empirical moments and $\widetilde{\mathbf{m}}_s(\boldsymbol{\theta}) \in \mathbb{R}^{p}$ the corresponding vector of simulated moments, MSM estimates parameters by solving
\begin{equation}
\widehat{\boldsymbol{\theta}} \in \arg\min_{\boldsymbol{\theta}} \left(\widetilde{\mathbf{m}}_s(\boldsymbol{\theta}) - \mathbf{m}_D\right)^\top W \left(\widetilde{\mathbf{m}}_s(\boldsymbol{\theta}) - \mathbf{m}_D\right),
\end{equation}
where $W$ is a symmetric, positive-definite weighting matrix that scales the contribution of each moment condition \cite{GrazziniRichiardi2015,Alahmadi2024}. The method is often used in dynamic and stochastic settings; moment conditions are typically chosen to be sensitive to the parameters of interest and to have reasonably low sampling variance \cite{GrazziniRichiardi2015}.

In this paper, we implement MSM in the standard simulated minimum distance form: it compares vectors of simulated and empirical moments and weights their squared differences \cite{Platt2020}. Let $\widehat{\mathbf{m}}$ denote the vector of empirical moments (e.g., variance, kurtosis, and selected autocorrelations of the raw, absolute, and squared series) and let $\widehat{\mathbf{m}}^{s}_i(\boldsymbol{\theta})$ be the corresponding moment vector computed from the $i$th Monte Carlo replication $X^{s}_i(\boldsymbol{\theta}, T_{\text{sim}})$, $i = 1,\dots,R$. Following the MSM implementation in our code, we use a seven-dimensional moment vector
\begin{equation}
\widehat{\mathbf{m}}^{s}_i(\boldsymbol{\theta}) = \bigl(\operatorname{Var}(x), \operatorname{Kurt}(x), \rho_x(1), \rho_{|x|}(1), \rho_{x^2}(1), \rho_{|x|}(5), \rho_{x^2}(5)\bigr)^\top.
\end{equation}
We define the average moment discrepancy
\begin{equation}
g(\boldsymbol{\theta}) = \frac{1}{R} \sum_{i=1}^R \left( \widehat{\mathbf{m}}^{s}_i(\boldsymbol{\theta}) - \widehat{\mathbf{m}} \right),
\end{equation}
and the MSM objective
\begin{equation}
f_{\text{MSM}}(\boldsymbol{\theta}) = g(\boldsymbol{\theta})^\top W g(\boldsymbol{\theta}),
\end{equation}
where $W$ is a $7\times 7$ weighting matrix given by the inverse of the Newey--West estimator of the covariance matrix of the empirical moments (with a small regularisation on the diagonal, as in our implementation). Minimizing $f_{\text{MSM}}$ corresponds to a generalized least-squares fit between simulated and empirical moment vectors. In more intuitive terms, MSM chooses parameter values so that simple summary statistics of the simulated ABM, such as its volatility, tail thickness, and autocorrelation patterns, match the corresponding statistics computed from real data as closely as possible \cite{Alahmadi2024}.

\subsection{Markov Information Criteria (MIC)}\label{app:mic}

In the MIC framework \cite{Barde2017}, a \emph{prediction} is a conditional probability mass function over the discrete states the system can occupy, given its history, and a \emph{model} is any device that produces a complete set of such predictions. The Markov Information Criteria (MIC) was introduced to compare the explanatory power of different models using simulated data \cite{Alahmadi2024,Barde2017}. It estimates the Kullback--Leibler (KL) distance between the model's predictive distribution and the true data-generating distribution, using the data compression interpretation of the KL distance and the minimum description length (MDL) principle \cite{Barde2017}. The context tree weighting (CTW) algorithm \cite{Barde2017} forms the basis for MIC, giving optimal cross-entropy measurement and universality over Markov processes of arbitrary order. MIC was applied to ABM calibration by \cite{Platt2020}. Operationally, it compares simulated and empirical series via cross-entropy without requiring a fixed set of moments \cite{Alahmadi2024}. Each observation is first discretised into $r$ binary digits, yielding a sequence $\tilde{X}_t \in \{0,1\}^r$. A context tree weighting (CTW) model (or, in our implementation, an $L$-order Markov model with Krichevsky--Trofimov smoothing) is then fitted to the ensemble of simulated series to obtain, for each time $t$ and bit $k$, a probability $\hat{p}^{\theta}_t(k)$ that $\tilde{X}_t^{\{k\}} = 1$ given the previous $L$ discretised observations. A second model is fitted to the empirical series to obtain probabilities $\hat{p}^{\text{emp}}_t(k)$.

For each time $t$ we define the simulated-model log-loss (cross-entropy) and empirical-model log-loss (entropy)
\begin{align}
\lambda_t(\boldsymbol{\theta}) &= - \sum_{k=1}^r \Bigl[\tilde{X}_t^{\{k\}} \log_2 \hat{p}^{\theta}_t(k) + \bigl(1 - \tilde{X}_t^{\{k\}}\bigr) \log_2 \bigl(1 - \hat{p}^{\theta}_t(k)\bigr)\Bigr],\\
\varepsilon_t &= - \sum_{k=1}^r \Bigl[\tilde{X}_t^{\{k\}} \log_2 \hat{p}^{\text{emp}}_t(k) + \bigl(1 - \tilde{X}_t^{\{k\}}\bigr) \log_2 \bigl(1 - \hat{p}^{\text{emp}}_t(k)\bigr)\Bigr].
\end{align}
The MIC objective is then
\begin{equation}
f_{\text{MIC}}(\boldsymbol{\theta}) = \sum_{t=L+1}^{T_{\text{emp}}} \bigl[\lambda_t(\boldsymbol{\theta}) - \varepsilon_t\bigr],
\end{equation}
which is an estimate of the Kullback--Leibler divergence between the simulated and empirical distributions of discretised series. Lower values indicate better agreement between model and data. Informally, MIC asks how much worse a compact Markov model trained on simulated data is at predicting (or compressing) the empirical series than a model trained directly on the empirical series; parameters are preferred when this prediction gap is small \cite{Alahmadi2024}.

\subsection{Generalized subtracted $L$-divergence (GSL-div)}\label{app:gsl}

The generalized subtracted $L$-divergence (GSL-div) is an information-theoretic objective introduced in \cite{Lamperti2016} to measure distance between time series data and to assess the empirical validity of simulated models \cite{Alahmadi2024}. In this framework, ``real-world data'' are the empirically observable elements of the system, and ``empirical validity'' is the degree of similarity between a model's output and real-world data; GSL-div can be used to quantify that validity and to discriminate between competing models \cite{Lamperti2016}. Put simply, GSL-div turns both time series into sequences of symbols, counts how often short patterns of symbols occur at different window lengths, and measures how different those pattern distributions are; thus it compares model and data directly in the time domain without choosing a particular set of moments.

Formally, GSL-div builds on the Kullback--Leibler (KL) divergence between two discrete distributions $p$ and $q$ on a common support $S$,
\begin{equation}
D_{\mathrm{KL}}(p \| q) = \sum_{s \in S} p(s) \log\frac{p(s)}{q(s)},
\end{equation}
and on the symmetric $L$-divergence \cite{Lamperti2016}
\begin{equation}
D_L(p \| q) = D_{\mathrm{KL}}(p \| m) + D_{\mathrm{KL}}(q \| m), \quad m = \frac{p+q}{2},
\end{equation}
which compares the distributions of time-changes in real-world data and in model-generated output. GSL-div estimates a subtracted version of this $L$-divergence at multiple time scales and aggregates the results into a single criterion that is bounded and well-defined for discrete symbolised series \cite{Lamperti2016,Alahmadi2024}. In our implementation, both empirical and simulated series are first discretised into an alphabet of $b$ symbols. For each window length $l = 1,\dots,L$ we consider the vocabulary $S_{l,b}$ of all possible words (length-$l$ sequences of symbols) and estimate
\begin{itemize}
\item $\hat{p}_l(s)$: empirical relative frequency of word $s \in S_{l,b}$,
\item $\hat{p}^{\theta}_l(s)$: average simulated relative frequency of $s$ across Monte Carlo replications,
\item $m^{\theta}_l(s) = \bigl(\hat{p}_l(s) + \hat{p}^{\theta}_l(s)\bigr)/2$: the midpoint distribution.
\end{itemize}
Following the ``subtracted $L$-divergence'' of \cite{Lamperti2016}, we define for each $l$ the divergence term
\begin{equation}
\mathcal{D}_l(\boldsymbol{\theta}) = -2 \sum_{s \in S_{l,b}} m^{\theta}_l(s)\log m^{\theta}_l(s)
 + \sum_{s \in S_{l,b}} \hat{p}_l(s)\log \hat{p}_l(s)
 + \sum_{s \in S_{l,b}} \hat{p}^{\theta}_l(s)\log \hat{p}^{\theta}_l(s),
\end{equation}
and a small-sample bias correction
\begin{equation}
\text{Bias}_l(\boldsymbol{\theta}) = \frac{B_{m^{\theta}_l} - 1}{4\,N_l} - \frac{B_{\hat{p}^{\theta}_l} - 1}{2\,N_l},
\end{equation}
where $N_l = T_{\text{sim}} - l + 1$ is the number of windows in a single series, $B_{m^{\theta}_l}$ is the number of words with $m^{\theta}_l(s) > 0$, and $B_{\hat{p}^{\theta}_l}$ is the number of words with $\hat{p}^{\theta}_l(s) > 0$. The GSL-div objective aggregates these contributions over window lengths:
\begin{equation}
f_{\text{GSL}}(\boldsymbol{\theta}) = \sum_{l=1}^L w_l \Bigl[\mathcal{D}_l(\boldsymbol{\theta}) + \text{Bias}_l(\boldsymbol{\theta})\Bigr],
\end{equation}
with non-negative weights $w_l$ such that $\sum_l w_l = 1$ (in our experiments, uniform weights). Smaller values indicate closer agreement between empirical and simulated word distributions across time scales \cite{Alahmadi2024}.

\subsection{Brock--Hommes (BH) model}\label{app:bh}

In their seminal contribution, Brock and Hommes~\cite{BrockHommes1998} develop an asset pricing model (referred to here as B\&H) in which a heterogeneous population of agents trade a generic asset according to $H$ different strategies (fundamentalist, chartist, etc.).
The presentation below follows~\cite{Lamperti2018,Platt2020}.

There is a population of $N$ traders that can invest either in a risk-free asset, which is perfectly elastically supplied at a gross return $R = 1 + r > 1$, or in a risky asset that pays an uncertain dividend~$y$ and has a price denoted by~$p$.
Wealth dynamics are given by
\begin{equation}\label{eq:bh-wealth}
  W_{t+1} = R\,W_t + (p_{t+1} + y_{t+1} - R\,p_t)\,z_t,
\end{equation}
where $p_{t+1}$ and $y_{t+1}$ are random variables and $z_t$ is the number of shares of the risky asset bought at time~$t$.

Traders are heterogeneous in their expectations about future prices and dividends and are assumed to be myopic mean-variance maximizers.
Because past prices and dividends are publicly available, agents can form conditional expected values $E_t$ and variances $V_t$.
The demand $z_{h,t}$ of agents with expectations of type~$h$ is obtained by solving
\begin{equation}\label{eq:bh-maxutil}
  \max_{z_{h,t}} \Bigl\{ E_{h,t}(W_{t+1}) - \tfrac{\nu}{2}\,V_{h,t}(W_{t+1}) \Bigr\},
\end{equation}
which yields
\begin{equation}\label{eq:bh-demand}
  z_{h,t} = \frac{E_{h,t}(p_{t+1} + y_{t+1} - R\,p_t)}{\nu\,\sigma^2},
\end{equation}
where $\nu$ controls risk aversion and $\sigma$ denotes conditional volatility, assumed equal across traders and constant over time.

With zero net supply of outside shares and $H$ trader types, market equilibrium requires
\begin{equation}\label{eq:bh-equil}
  R\,p_t = \sum_{h=1}^{H} n_{h,t}\,E_{h,t}(p_{t+1} + y_{t+1}),
\end{equation}
where $n_{h,t}$ is the fraction of traders following strategy~$h$ at time~$t$.
Under homogeneous traders, perfect information, and rational expectations, the no-arbitrage equilibrium condition is
\begin{equation}\label{eq:bh-noarb}
  R\,p_t^{*} = E_t(p_{t+1}^{*} + y_{t+1}),
\end{equation}
where $p^{*}$ denotes the fundamental price.
When dividends are i.i.d.\ with constant mean, the fundamental price is $p^{*} = E(y_t)/(R-1)$.
Prices are henceforth expressed as deviations from the fundamental, i.e.\ $x_t = p_t - p_t^{*}$.

At the beginning of each trading period $t = 1, 2, \ldots, T$, agents form heterogeneous forecasts.
Specifically, investors believe that prices may deviate from the fundamental value according to some function $f_h(\cdot)$ of past deviations.
The beliefs of agents of type~$h$ evolve as
\begin{equation}\label{eq:bh-beliefs}
  E_{h,t}(p_{t+1} + y_{t+1}) = E_t(p_{t+1}^{*}) + f_h(x_{t-1}, \ldots, x_{t-L}).
\end{equation}
In the specification adopted here, each strategy~$h$ uses a linear belief rule
\begin{equation}\label{eq:bh-linear-belief}
  f_h(x_{t-1}) = g_h\,x_{t-1} + b_h,
\end{equation}
where $g_h$ is a trend-following (or contrarian) parameter and $b_h$ is a bias term.

Substituting Eqs.~\eqref{eq:bh-beliefs}--\eqref{eq:bh-linear-belief} into the equilibrium condition~\eqref{eq:bh-equil} and adding a noise term yields the closed-form price dynamics in deviations\footnote{The interested reader should refer to~\cite{BrockHommes1998} for a detailed discussion of the model's underlying assumptions and the derivation of the closed-form solution.}:
\begin{equation}\label{eq:bh-deviation-equil}
  x_{t+1} = \frac{1}{1+r}\left[\sum_{h=1}^{H} n_{h,t+1}\bigl(g_h\,x_t + b_h\bigr) + \varepsilon_{t+1}\right],
\end{equation}
where $\varepsilon_t \sim \mathcal{N}(0, \sigma^2)$ and $R = 1 + r$.
Traders switch among strategies according to their evolving profitability.
The probability that an agent adopts strategy~$h$ is given by the multinomial logit
\begin{equation}\label{eq:bh-switching}
  n_{h,t+1} = \frac{\exp(\beta\,U_{h,t})}{\sum_{j=1}^{H} \exp(\beta\,U_{j,t})},
\end{equation}
where $\beta \in [0, +\infty)$ captures the intensity of choice: higher $\beta$ means agents switch more aggressively toward profitable strategies, while the stochastic element ensures the system never collapses to a single-strategy equilibrium.
Each strategy~$h$ is associated with a fitness measure
\begin{equation}\label{eq:bh-fitness}
  U_{h,t} = (x_t - R\,x_{t-1})(g_h\,x_{t-2} + b_h - R\,x_{t-1}),
\end{equation}
which captures the realized profit of strategy~$h$ based on past price deviations.

\subsection{Island growth model}\label{app:island}

The Island growth model, introduced by~\cite{FagioloDosi2003}, describes an economy in which heterogeneous firms explore and exploit technological opportunities distributed across a discrete spatial lattice.
The model is used as a second calibration benchmark in our experiments and is substantially more complex than the B\&H model, featuring seven free parameters and richer emergent dynamics~\cite{Lamperti2018}.

The economy is represented as a two-dimensional lattice of cells.
Each cell may contain an \emph{island}, i.e.\ a production opportunity with an associated productivity level.
At the start of the simulation, each cell is independently designated as an island with probability~$\pi$.
A population of~$m_0$ firms begins on a single ``home'' island and, at each time step, each firm occupies one of three behavioural modes: \emph{mining} (exploitation), \emph{exploring}, or \emph{imitating}.

A miner extracts output from its current island.
The aggregate production of island~$j$ at time~$t$ is
\begin{equation}\label{eq:island-production}
  Q_{j,t} = s_{j,t}\,m_{j,t}^{\alpha},
\end{equation}
where $s_{j,t}$ is the island's productivity, $m_{j,t}$ is the number of miners currently on island~$j$, and $\alpha > 0$ controls returns to labour.
A miner switches to exploration with probability~$\varepsilon$ at each step: it then performs a random walk on the lattice until it discovers an unoccupied island, which it colonises.

The productivity of a newly discovered island~$j$ depends on the discoverer's accumulated experience through a cumulative learning mechanism:
\begin{equation}\label{eq:island-productivity}
  s_{j,0} = 1 + \varphi\,Q_i^{\mathrm{cum}},
\end{equation}
where $Q_i^{\mathrm{cum}}$ is the cumulative output of firm~$i$ up to the time of discovery and $\varphi \in [0,1]$ governs the strength of the learning effect.
This mechanism captures the notion that experienced firms are better at exploiting new technological paradigms.

After a transient period, miners may also become imitators.
An imitator selects a target island based on a knowledge-diffusion signal that decays with distance: the probability that firm~$i$ on island~$k$ receives a signal from island~$j$ is
\begin{equation}\label{eq:island-signal}
  \Pr(\text{signal from } j \mid i \text{ on } k) = \frac{s_{j,t}\,m_{j,t}^{\alpha}}{\bigl(1 + d(k,j)\bigr)^{\rho}},
\end{equation}
where $d(k,j)$ is the Manhattan distance between islands~$k$ and~$j$, and $\rho \geq 0$ controls the locality of knowledge diffusion; higher values make distant islands less visible.
On arrival at the target island, the imitator resumes mining, so imitation provides a mechanism for technology diffusion across the lattice.

Additionally, islands may experience random productivity jumps: at each time step, each island's productivity is incremented by a Poisson-distributed shock with mean~$\lambda$, capturing exogenous technological breakthroughs.

The model's aggregate output is GDP, computed as the sum of island-level production across all active islands at each time step.
In our experiments the simulation runs for $T = 1000$ periods.
The seven free parameters $\boldsymbol{\theta} = (\rho, \alpha, \varphi, \pi, \varepsilon, m_0, \lambda)$ are the calibration targets; their true values and search bounds are listed in Table~\ref{tbl:island-params}.
Compared with the four-parameter B\&H model, the Island model's higher dimensionality and richer dynamics make it a more demanding test of calibration performance.

\section{Supplementary Tables}\label{app:tables}

\begin{table}[H]
\centering
\footnotesize
\setlength{\tabcolsep}{4pt}
\caption{Brock--Hommes model: pure-optimizer vs.\ best surrogate-assisted comparison for each optimizer--objective pair. The best surrogate is the one with the lowest mean RMSE across five independent runs; positive improvement means it reduces error or computation time relative to the pure baseline, negative that it is worse. \rev{Accuracy improvement is negative for all four GA pairs and positive for all four PSO pairs, the split also visible in \mbox{Figure~\ref{fig:bh-radar-rmse}}, whereas the time improvement is between \mbox{$34.4\%$} and \mbox{$36.6\%$} for every pair.}}
\label{tbl:bh-improvement-accuracy}
\begin{tabular}{@{}lll rr r rr r@{}}
\toprule
 & & & \multicolumn{3}{c}{Accuracy (RMSE)} & \multicolumn{3}{c}{Speed (seconds)} \\
\cmidrule(lr){4-6} \cmidrule(l){7-9}
Optimizer & Objective & Best Surrogate & Pure & Surrogate & Impr.\ (\%) & Pure & Surrogate & Impr.\ (\%) \\
\midrule
GA  & GSL-div & \rev{ANN}           & \rev{0.4497} & \rev{0.6086} & \rev{\mbox{$-$35.3}} & \rev{8795.2} & \rev{5724.7} & \rev{\mbox{$+$34.9}} \\
GA  & MSM     & \rev{ANN}           & \rev{0.3251} & \rev{0.3761} & \rev{\mbox{$-$15.7}} & \rev{8734.3} & \rev{5549.3} & \rev{\mbox{$+$36.5}} \\
GA  & MIC     & \rev{ANN}           & \rev{0.2856} & \rev{0.3985} & \rev{\mbox{$-$39.5}} & \rev{8739.1} & \rev{5570.0} & \rev{\mbox{$+$36.3}} \\
GA  & KS-test & \rev{ANN}           & \rev{0.8513} & \rev{0.8682} & \rev{\mbox{$-$2.0}} & \rev{8860.5} & \rev{5740.8} & \rev{\mbox{$+$35.2}} \\
PSO & GSL-div & \rev{ANN}           & \rev{0.4814} & \rev{0.2758} & \rev{\mbox{$+$42.7}} & \rev{9215.0} & \rev{5939.6} & \rev{\mbox{$+$35.5}} \\
PSO & MSM     & \rev{Random Forest} & \rev{0.6863} & \rev{0.3780} & \rev{\mbox{$+$44.9}} & \rev{9152.1} & \rev{5806.6} & \rev{\mbox{$+$36.6}} \\
PSO & MIC     & \rev{ANN}           & \rev{0.4212} & \rev{0.2286} & \rev{\mbox{$+$45.7}} & \rev{9046.1} & \rev{5932.4} & \rev{\mbox{$+$34.4}} \\
PSO & KS-test & \rev{Random Forest} & \rev{1.0170} & \rev{0.8244} & \rev{\mbox{$+$18.9}} & \rev{9190.6} & \rev{5916.9} & \rev{\mbox{$+$35.6}} \\
\bottomrule
\end{tabular}
\end{table}

\begin{table}[H]
\centering
\footnotesize
\setlength{\tabcolsep}{4pt}
\caption{Island growth model: pure-optimizer vs.\ best surrogate-assisted comparison, with columns and conventions as in \mbox{Table~\ref{tbl:bh-improvement-accuracy}}. \rev{In contrast to the BH pattern, accuracy improvement is non-negative for all eight pairs, from \mbox{$0.0\%$} on the two MSM pairs, where no surrogate alters the result at all, up to \mbox{$65.8\%$}. Time improvement is far more variable than the uniform \mbox{$34$}--\mbox{$37\%$} seen on BH, spanning \mbox{$-6.2\%$} to \mbox{$+70.4\%$} and negative only on those same two MSM pairs.}}
\label{tbl:island-improvement-accuracy}
\begin{tabular}{@{}lll rr r rr r@{}}
\toprule
 & & & \multicolumn{3}{c}{Accuracy (RMSE)} & \multicolumn{3}{c}{Speed (seconds)} \\
\cmidrule(lr){4-6} \cmidrule(l){7-9}
Optimizer & Objective & Best Surrogate & Pure & Surrogate & Impr.\ (\%) & Pure & Surrogate & Impr.\ (\%) \\
\midrule
GA  & GSL-div & \rev{ANN}           & \rev{8.7008} & \rev{3.3922} & \rev{\mbox{$+$61.0}} & \rev{30233.4} & \rev{11614.9} & \rev{\mbox{$+$61.6}} \\
GA  & MSM     & \rev{ANN}           & \rev{6.5874} & \rev{6.5874} & \rev{\mbox{$+$0.0}} & \rev{23899.5} & \rev{24400.4} & \rev{\mbox{$-$2.1}} \\
GA  & MIC     & \rev{ANN}           & \rev{10.3711} & \rev{7.4523} & \rev{\mbox{$+$28.1}} & \rev{27347.6} & \rev{8097.9} & \rev{\mbox{$+$70.4}} \\
GA  & KS-test & \rev{ANN}           & \rev{6.9732} & \rev{2.3862} & \rev{\mbox{$+$65.8}} & \rev{23440.8} & \rev{9307.5} & \rev{\mbox{$+$60.3}} \\
PSO & GSL-div & \rev{LR}            & \rev{13.5182} & \rev{10.3046} & \rev{\mbox{$+$23.8}} & \rev{27237.2} & \rev{17628.1} & \rev{\mbox{$+$35.3}} \\
PSO & MSM     & \rev{ANN}           & \rev{18.9118} & \rev{18.9118} & \rev{\mbox{$+$0.0}} & \rev{8150.3} & \rev{8656.6} & \rev{\mbox{$-$6.2}} \\
PSO & MIC     & \rev{ANN}           & \rev{12.7161} & \rev{7.3569} & \rev{\mbox{$+$42.1}} & \rev{7210.0} & \rev{4290.3} & \rev{\mbox{$+$40.5}} \\
PSO & KS-test & \rev{Random Forest} & \rev{13.6530} & \rev{5.9595} & \rev{\mbox{$+$56.4}} & \rev{10142.1} & \rev{8645.0} & \rev{\mbox{$+$14.8}} \\
\bottomrule
\end{tabular}
\end{table}

\FloatBarrier

\bibliographystyle{unsrt}
\bibliography{cas-refs}

\end{document}